\documentclass[runningheads]{llncs}

\usepackage{eccv}

\usepackage{eccvabbrv}
\usepackage{graphicx}
\usepackage{booktabs}
\usepackage[accsupp]{axessibility}  
\usepackage{kotex}
\usepackage{multirow}
\usepackage{algorithm}
\usepackage{algpseudocode}
\usepackage{amsmath}
\usepackage{amssymb}
\usepackage{graphicx}
\usepackage{wrapfig}
\newcommand{\tworow}[2]{\begin{tabular}[c]{@{}c@{}}#1\vspace{-2pt}\\#2\end{tabular}}
\usepackage{setspace}

\usepackage{booktabs}
\usepackage[table]{xcolor}
\usepackage{wrapfig}
\usepackage{animate}
\usepackage[export]{adjustbox}
\usepackage{tikz}
\usepackage{float}
\usepackage{cuted}
\usepackage[hypcap=false]{caption}
\usepackage[percent]{overpic}

\usepackage{orcidlink}

\makeatletter
\renewcommand\@fnsymbol[1]{\ensuremath{\ifcase#1\or\dagger\or\ddagger\or
  \mathsection\or\mathparagraph\fi}}
\makeatother

\begin{document}

\title{PIAvatar: Physically Interactive Avatars via Deformation Gradient Decoupling} 

\titlerunning{PIAvatar}


\author{
Sang-Hun Han\inst{1}\orcidlink{0009-0001-8019-265X} \and
Min-Gyu Park\inst{2,3}\orcidlink{0000-0003-1752-150X} \and
Jisu Shin\inst{1}\orcidlink{0009-0007-7640-1512} \and \\
Seunghyun Shin\inst{1}\orcidlink{0009-0006-3012-9675} \and 
Jin-Hwi Park\inst{4}\orcidlink{0000-0001-7874-2344} \and
Hae-Gon Jeon\inst{5}\thanks{Corresponding author.}\orcidlink{0000-0003-1105-1666}
}

\authorrunning{S. Han et al.}

\institute{$^{1}$GIST, $^{2}$KETI, $^{3}$Polygom, $^{4}$Chung-Ang University, \\
$^{5}$Department of Artificial Intelligence, Yonsei University\\
\email{\tt\small \{sanghunhan, jsshin98, seunghyuns98\}@gm.gist.ac.kr, \\
\tt\small mpark@keti.re.kr,
\tt\small jinhwipark@cau.ac.kr,
\tt\small earboll@yonsei.ac.kr}}


\maketitle

\vspace{-20pt}
\begin{abstract}
3D human avatars have shown impressive visual fidelity \allowbreak driven by pose-conditioned models, yet they still lack the physical ability required for interactions with each other and environments. 
Although recent studies have made various attempts to incorporate physical characteristics into 3D avatars, they only exhibit limited physical deformations, often leading to constrained interaction behaviors.
To resolve this issue, we present PIAvatar, a framework to simultaneously enable physically aware interactions between avatar-avatar and avatar-environment, and a non-rigid deformable human body simulation. 
In this work, our key insight is to decouple kinematic velocity from deformation gradient. When external forces act on avatars, the kinematic velocity induces stress which hinders the avatar's ability to achieve a desired pose.
In addition, we integrate a skeletal framework within the avatar. It allows estimating its poses and real-time tracking in a closed form, even during non-rigid physical interactions.
Our approach is implemented within a conventional Material Point Method framework to ensure physically consistent dynamics.
We lastly evaluate the method on both human-object and human-human interaction scenarios to assess its behavior under diverse interaction settings.
\end{abstract}
\vspace{-15pt}

\begin{figure*}[t]
    \centering
    \captionsetup{type=figure}
    \includegraphics[width=\linewidth]{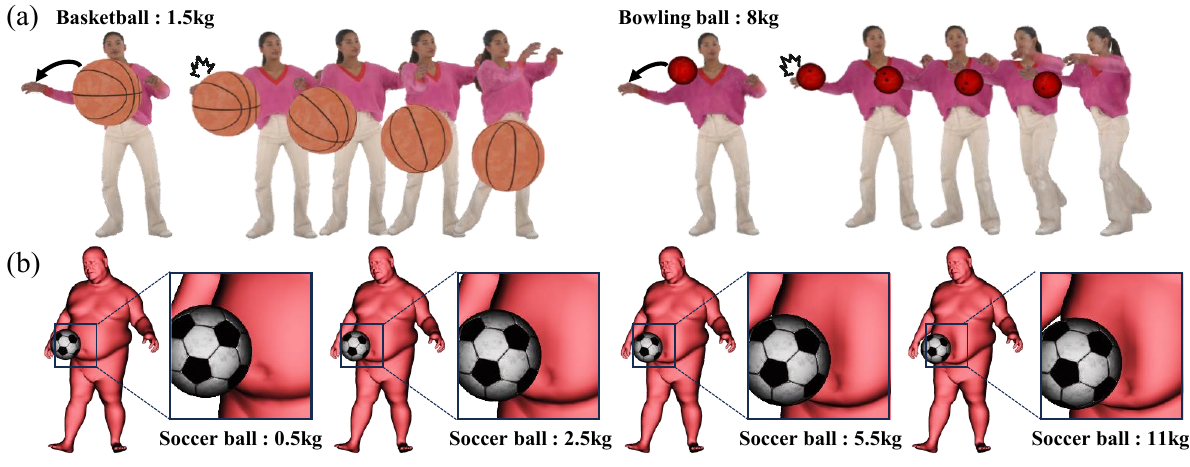}
    \vspace{-17pt}
    \captionof{figure}{\textbf{Physical avatar interactions with bidirectional and non-rigid deformations.} PIAvatar enables bidirectional interactions between avatars and their surroundings within a unified simulation framework. (a) Varying the mass of the ball changes both the avatar pose and the ball trajectory, illustrating bidirectional human–object interaction. (b) The avatar exhibits different non-rigid deformations depending on the mass of the soccer ball during impact.}
    \vspace{-15pt}
    \label{fig:teaser}
\end{figure*}

\section{Introduction}
\label{sec:intro}

The generation of realistic 3D human avatars is crucial for bridging virtual and real-world experiences, enabling lifelike interactions in diverse AR/VR applications. 
Prior research has primarily focused on detailed human geometry and appearance, leveraging various representations such as implicit and volumetric models, 3D Gaussians~\cite{kerbl20233d}, and parametric body models~\cite{loper2015smpl, SMPLX_2019_Pavlakos, anguelov2005scape, xu2020ghum}. 
These approaches have shown promising visual fidelity, producing realistic renderings across a wide range of body configurations. 
However, since physical properties are not built in these models, we can access only kinematic animations of avatars with no physical interactions.
That is, such avatars remain fundamentally incapable of interacting with each other and their surrounding environments. 

Several studies have recently begun to explore physically aware avatars.
Reinforcement learning (RL)–based approaches~\cite{tevet2024closd, xu2025intermimic} allow avatars to interact with objects. Due to their large-scale parallel training setting, they usually rely on simplified physics environments~\cite{2018-TOG-deepMimic, coumans2015bullet, nvidiaPhysX, todorov2012mujoco}. 
This constraint limits their ability to represent non-rigid deformations limiting their expressiveness to simplified geometric forms such as cylinders or other primitive shapes.
Meanwhile, simulation-driven methods, such as dressed avatar physics~\cite{zheng2024physavatar, lee2025mpmavatar} or kinematic models~\cite{siyao2025half} augmented with partial dynamics, take promising steps toward physical realism. 
They either allow one-way interactions, where the avatar can exert forces on surrounding objects but not vice versa, or fail to simulate physical surface deformations of the avatar. For example, Half-Physics~\cite{siyao2025half} allows external forces to influence the avatar, but its responses remain at the pose level without modeling surface deformation.

To address these limitations, we propose PIAvatar, a framework for physically consistent human avatar interactions. 
As illustrated in Fig.~\ref{fig:teaser}, our method is developed with two primary objectives:  
(1) \textit{Bidirectional physical interactions} including avatar-avatar and avatar-object; and (2) \textit{Non-rigid deformation} to undergo natural interactions under physical forces.
To achieve this, we build our simulation system based on the Material Point Method (MPM), a particle-based continuum simulation framework. Even though MPM inherently supports non-rigid deformation and bidirectional momentum transfer,
it introduces two key challenges when driving an avatar with user-defined motions. 
First, all objects inside MPM, modeled with a stress-based constitutive model, naturally induce restorative forces when deformed. The restorative forces counteract the externally applied kinematic velocity and hinder avatars from reaching the intended target pose. 
Second, once the avatar undergoes non-rigid deformation through environmental interactions, its pose is no longer directly trackable. 
While forward animation can easily generate deformations from a given pose, the inverse process—recovering pose from a physically deformed shape—is non-trivial and often ill-posed.
Consequently, it is typically solved through computationally expensive methods such as parametric model fitting~\cite{loper2015smpl, SMPLX_2019_Pavlakos, anguelov2005scape}, non-linear optimization~\cite{Bogo:ECCV:2016, loper2014mosh}, or learning–based regression~\cite{hmrKanazawa17, kocabas2019vibe}.

To overcome these challenges, we propose two approaches. We explicitly exclude a user-defined kinematic velocity from a deformation gradient computation. This prevents undesired stress and enables the avatar to follow the intended motion without the restorative resistance.
We then incorporate the deformation process into the MPM framework. 
In addition, we embed a skeletal structure inside an avatar and compute its pose in a closed form, enabling robust and direct pose estimation even under non-rigid physical interactions. With this integrated design, our system enables physically consistent and controllable interactions between avatars and their surrounding environments.

{\sloppy
We demonstrate the capabilities of PIAvatar through various avatar–\allowbreak environment interaction scenarios, including both avatar-\allowbreak object and avatar-\allowbreak avatar contacts. 
Our avatars exhibit physical, non-rigid body deformations under collisions and external forces. 
We also show that varying material properties, such as density, produce diverse and physically consistent simulation outcomes.
\par}

In summary, our contributions are as follows:
\begin{enumerate}
\item We present PIAvatar, a novel MPM-based avatar simulation framework that enables both bidirectional physical interactions and non-rigid deformations.
\item For this, we decouple the deformation gradient to ensure the user-defined kinematic velocity directly drives the avatar’s motion without inducing stress while still allowing physical interactions, and embed a skeletal structure for direct pose estimation. Note that these whole processes are solved in a closed form optimization, not in a learning-based manner.
\item We demonstrate physically grounded interactions across multiple human–\allowbreak object and human–human scenarios using PIAvatar. 
\end{enumerate}
\vspace{-10pt}

\section{Related Work}
\label{sec:related}
\vspace{-5pt}

\subsection{Animatable Human Avatar Generation}
Building on parametric human body models~\cite{loper2015smpl, pavlakos2019expressive}, numerous studies have explored the reconstruction of posed human avatars directly from single images~\cite{saito2019pifu, saito2020pifuhd, zheng2021pamir, xiu2022icon, xiu2023econ, han2023high, ho2024sith, xue2024human}. 
However, those works typically focus on static or per-frame reconstruction without explicit animation capability.
To achieve animatable avatars, subsequent works leverage Linear Blend Skinning (LBS)-based deformation to integrate geometric or texture information from multiple images into a canonical space~\cite{peng2021neural, weng2022humannerf, su2021nerf, peng2021animatable, guo2023vid2avatar, jiang2023instantavatar, shin2024canonicalfusion}.
Recently, 3D or 4D Gaussian Splatting (GS)~\cite{kerbl20233d, wu20244d} is incorporated into this framework, offering efficient rendering and high-quality geometry reconstruction~\cite{li2024animatable, hu2024gaussianavatar, wen2024gomavatar, pan2024humansplat, moon2024expressive, shao2024splattingavatar, hu2024expressive, qiu2025anigs}.
With these advances, recent approaches~\cite{wang2025fresa, zhuang2025idol, sim2025persona, qiu2025lhm} directly infer pose-dependent geometry and appearance without any per-scene optimization using large feed-forward models, allowing scalable and real-time animatable avatar generations.


Additionally, skeleton-based methods such as OSSO~\cite{keller2022osso} and SKEL~\cite{keller2023skin} have explored explicit skeletal representations to enhance articulation control and motion coherence. Such approaches demonstrate the growing integration of skeleton-driven modeling within animatable human representations, laying the groundwork for physically coupled simulation as pursued in our work.
\vspace{-10pt}


\subsection{Physics-based Avatar Simulation} 


From a perspective of physics-based methods, relevant studies can be broadly categorized into control-driven approaches and simulation-based deformable modeling.
Control-driven methods, including RL-based approaches~\cite{2018-TOG-deepMimic,tevet2024closd,xu2025intermimic,peng2021amp,luo2023perpetual,luo2023universal}, learn policies to generate physically plausible motions within physics simulators.
These methods focus on task-oriented control and reactive behaviors, typically relying on rigid-body or simplified physical representations for efficiency~\cite{todorov2012mujoco, coumans2015bullet, nvidiaPhysX}.
Simulation-based frameworks, including Finite Element Method (FEM)~\cite{clough1960finite}, C-IPC~\cite{li2020codimensional} and MPM~\cite{jiang2016material}, have tried to represent animatable human motions based on laws of physics such continuum mechanics.
Half-Physics~\cite{siyao2025half} couples a kinematic human model with a physics engine to enable pose-level adaptation under external forces.
C-IPC-based methods, which are mesh-based deformable simulation like PhysAvatar~\cite{zheng2024physavatar}, incorporate physics-inspired modeling to capture garment deformation effects.
FEM-based approaches~\cite{kim2017data,liu2013simulation}, explicitly model nonlinear soft-tissue dynamics using continuum mechanics formulations.
MPM-based approaches, such as MPMAvatar~\cite{lee2025mpmavatar}, use a particle–grid formulation to model volumetric deformation and simulate loose garments.
While these works advance physics-aware avatar modeling, achieving effective bidirectional interaction still remains challenging.
\vspace{-10pt}

\subsection{Material Point Method}

The MPM~\cite{jiang2016material} is a hybrid Eulerian-Lagrangian framework, introduced by Stomakhin \textit{et al.}~\cite{stomakhin2013material} for snow simulation, and has become a core technique for physically based animation of solids, fluids, and deformable bodies.
%
In the MPM, each particle carries a deformation gradient that couples the local velocity gradient with its deformation and stress.

To retain robustness, later studies such as APIC~\cite{jiang2015affine}, AMPIC~\cite{jiang2017angular}, and MLS-MPM~\cite{hu2018moving} enhanced momentum conservation and reduced numerical dissipation of kinetic energy. Other studies decomposed the deformation gradient into elastic and plastic parts for constitutive modeling \eg elasto-plastic snow~\cite{stomakhin2013material} and thin-shell rigidity corrections~\cite{guo2018material}.  
%
Further developments extended the capability of MPM into anisotropic, frictional, and hybrid soft-body materials~\cite{jiang2017anisotropic,klar2016drucker,han2019hybrid}. 
In parallel, Gaussian-based representations have been combined with physics-inspired modeling, such as PhysGaussian~\cite{xie2024physgaussian}, to enhance visual realism through a differentiable rendering and an appearance-driven regularization.
Overall, these advances have established the MPM as a versatile framework for physically grounded simulation. These properties make it a relevant foundation for modeling physically aware non-rigid human avatar interactions.

\section{Preliminary: Material Point Method}

\label{sec:preliminary_mpm}
In this work, we adopt the MPM framework to take advantage of its fast grid-based momentum exchange to efficiently model frequent interactions.
It also enables flexible deformations and complex contact interactions within a unified particle-grid framework.

To firstly explain a motion of a material space over time $t$, we begin with the conservation of mass and momentum as follows:
\begin{equation}
\label{eq:mass_momentum}
\frac{d\rho}{dt} + \rho \nabla \cdot \boldsymbol{v} = 0,
\qquad
\rho \frac{d\boldsymbol{v}}{dt} = \nabla \cdot \boldsymbol{\sigma} + \rho \boldsymbol{g},
\end{equation}
where $\rho$, $\boldsymbol{v}$, and $\boldsymbol{\sigma}$ denote the material density, velocity, and Cauchy stress, respectively. 
Following \cref{eq:mass_momentum}, continuum mechanics models the motion as a continuous deformation mapping $\boldsymbol{\phi}$, which takes spatial deformed coordinates $\boldsymbol{x} = \boldsymbol{\phi}(\boldsymbol{X}, t)$ from material coordinates $\boldsymbol{X}$. 
%
%
Its deformation gradient $\boldsymbol{F} = \partial \boldsymbol{\phi} / \partial \boldsymbol{X}$ encodes local transformations of the material, including rotation, stretch, and shear. 
Numerically solving these continuum equations in a purely Lagrangian form is computationally demanding under large deformations and frequent contact, which motivates the hybrid particle-grid formulation of MPM.

%
To model continuum mechanics, the Material Point Method (MPM) discretizes the continuum into Lagrangian material particles that track mass, momentum and deformation. Each particle $p$ is represented by its position $\boldsymbol{x}_p$, velocity $\boldsymbol{v}_p$, mass $m_p$, and deformation gradient $\boldsymbol{F}_p$.
%
An Eulerian grid solves the momentum equations and updates particle states through particle-grid transfers, where particle stresses are accumulated onto grid nodes to compute internal forces. While each particle computes its stress independently, the Cauchy stress $\boldsymbol{\sigma}_p$ is obtained from its deformation gradient $\boldsymbol{F}_p$ and the strain energy density $\Psi(\cdot)$ as:
\begin{equation}
    \boldsymbol{\sigma}_p = \frac{1}{J_p} \boldsymbol{F}_p \,
    \frac{\partial \Psi(\boldsymbol{F}_p)}{\partial \boldsymbol{F}_p}^{\!\top}
    \; \text{s.t.} \; J_p = \det(\boldsymbol{F}_p).
    \label{eq:kirchhoff_stress}
\end{equation}
However, in \cref{eq:kirchhoff_stress}, a challenge arises since all types of deformations are encoded with only $\boldsymbol{F}_p$, including user-imposed or purely kinematic deformations. 
In this paper, our solution to this challenge is to disentangle $\boldsymbol{F}_p$ into the kinematic part and the elastic part, inspired by snow and thin-shell material simulations~\cite{stomakhin2013material, guo2018material}.

\section{Proposed Method} 

\begin{figure*}[t]
    \centering
    \includegraphics[clip=true, width=\linewidth]{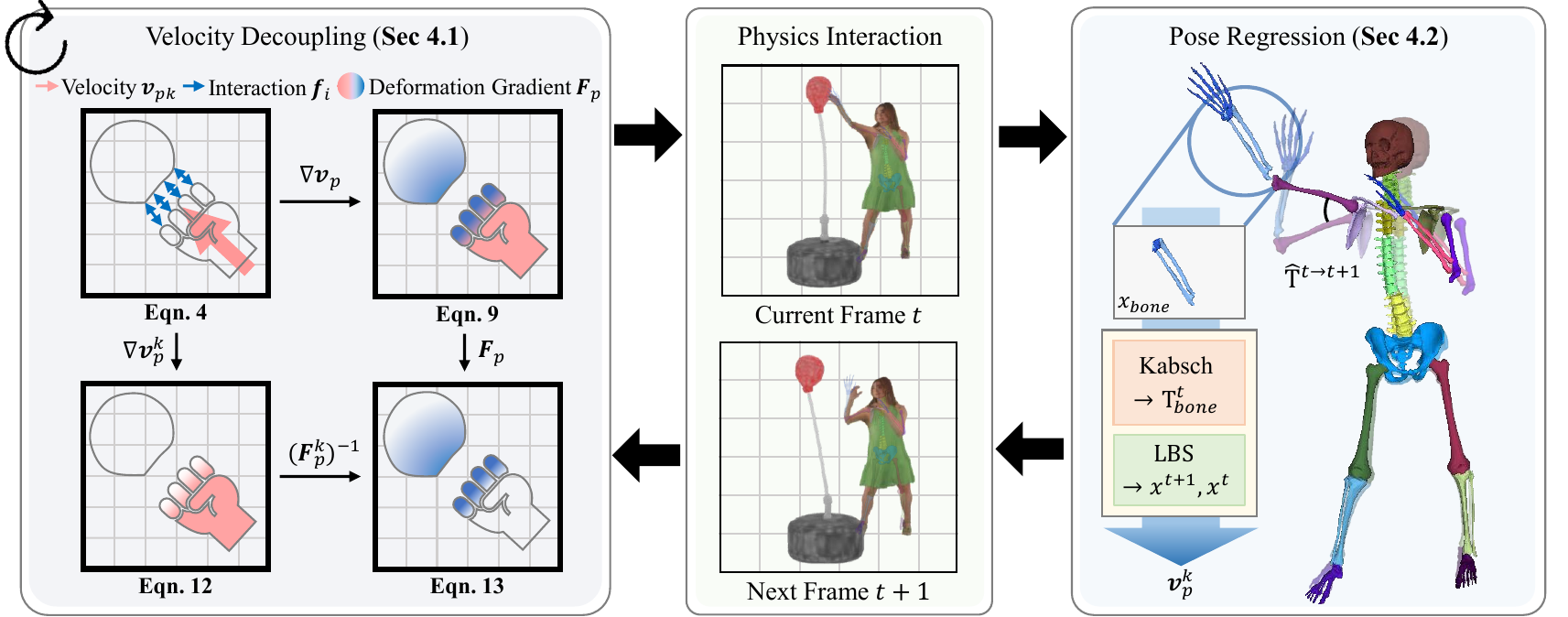}
    \vspace{-3mm}
    \caption{\textbf{An illustration of our framework.} (a) To faithfully reflect the user-defined motion, we decouple the kinematic velocity from the deformation gradient update (Sec.~\ref{sec:method1}). 
    (b) By computing the velocity from the transformations of the embedded skeletal structure, our method preserves the pose consistency throughout the simulation (Sec.~\ref{sec:method2}).}
    \vspace{-5mm}
    \label{fig:framework}
\end{figure*}

We first point out the problem in the basic MPM coming from stress-based forces, and present the solution that introduces the disentanglement of the user-defined kinematic velocity from the deformation gradient (Sec.~\ref{sec:method1}). We then introduce how to simulate physical avatar interactions with our skeletal pose extraction and kinematic velocity calculation (Sec.~\ref{sec:method2}). An overview of PIAvatar is illustrated in Fig.~\ref{fig:framework}.

\newpage

\subsection{Kinematic Deformation Decoupling} 
\vspace{-5mm}
\label{sec:method1}

\begin{wrapfigure}{r}{0.6\textwidth}
    \vspace{-8mm}
    \centering
    \includegraphics[clip=true, width=\linewidth]{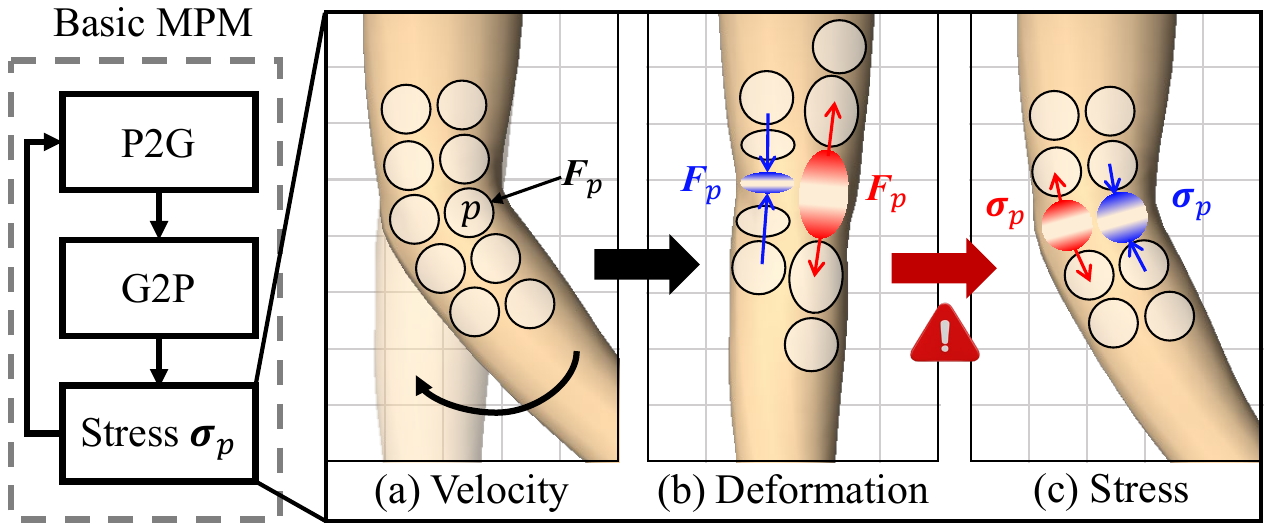}    
    \vspace{-6mm}
    \caption{\textbf{Velocity-induced stress formation.} (a) The kinematic velocity is applied to the avatar particles $p$. (b) The change in deformation gradients $\boldsymbol{F}_p$ occurs due to the kinematic velocity (see blue- and red-colored ellipses). (c) The deformation generates stress $\boldsymbol{\sigma}_p$ that hinders intended kinematic motion.}
    \label{fig:mpm}
    \vspace{-8mm}
\end{wrapfigure}


%
As shown in Fig.~\ref{fig:mpm}, we illustrate a common process of MPM. The particle-to-grid (P2G) and grid-to-particle (G2P) transfers allow us to compute physical quantities of a particle on a certain grid.
When the velocity is assigned to particles, they are interpolated to assign the velocity onto the grid during P2G.
As a result, the grid treats the velocity as physical momentum, which updates the deformation gradient $\boldsymbol{F}_p$ during G2P and produces the stress $\boldsymbol{\sigma}_p$. 

In particular, in non-rigid body motions like humans, the user-defined kinematic velocity is partially absorbed as the deformation gradient. This phenomenon results in the unintended stress that hinders accurate kinematic controls and leads to deviations from the desired pose. 
We point out this problem in a basic MPM with respect to the representational ability of user-preferred motions by describing its key components at first.


%

\noindent\textbf{Basic MPM.}\quad
In the P2G phase of a basic MPM, particle quantities, including position $\boldsymbol{x}_p$, velocity $\boldsymbol{v}_{p}$, mass $m_p$ and the Cauchy stress $\boldsymbol{\sigma}_p$, are interpolated to the grid using a weighting kernel $w_{ip}$:
\begin{align}
    m_i &= \sum_p w_{ip} \, m_p, \quad
    m_i\boldsymbol{v}_i =
    \sum_p w_{ip} \, m_p 
    \boldsymbol{v}_p
    + \Delta t \, \boldsymbol{f}_i, \label{eq:p2g_mv} \\ 
    \boldsymbol{f}_i &= \boldsymbol{f}_i^{\text{int}} + \boldsymbol{f}_i^{\text{ext}}\quad\text{s.t.}\quad
    \boldsymbol{f}_i^{\text{int}} = -\sum_p V_p \, \boldsymbol{\sigma}_p \, \nabla w_{ip}, \label{eq:internal_force}
\end{align}
where $i$ and $V_p$ denote the index of a grid node and the particle volume, respectively. Here, the weighting kernel $w_{ip}$ is the B-spline weight evaluated at the particle position relative to the $i$-th grid node.
The total force $\boldsymbol{f}_i$ on each grid is computed by summing external forces like gravity and the internal forces like stress that we want to handle. 

In the G2P phase, the velocity $\boldsymbol{v}_i$ on the grid is applied to the particle, updating the particle state such as $\boldsymbol{F}_p$:
\begin{align}
    \boldsymbol{v}_p &\leftarrow \sum_i w_{ip} \, \boldsymbol{v}_i, \quad\quad\quad
    \boldsymbol{x}_p \leftarrow \boldsymbol{x}_p + \Delta t \, \boldsymbol{v}_p,\\
    \nabla \boldsymbol{v}_p &= \sum_i \boldsymbol{v}_i (\nabla w_{ip})^\top,   \quad
    \boldsymbol{F}_p \leftarrow \left( \mathbf{I} +  \Delta t\nabla \boldsymbol{v}_p \right) \boldsymbol{F}_p.  \label{eq:dg_update}
\end{align}

In the MPM model, users can adjust $\boldsymbol{v}_p$ to control avatars, which contributes to the update of $\boldsymbol{F}_p$ in \cref{eq:dg_update}. According to \cref{eq:kirchhoff_stress}, the stress is obviously generated, introducing  $\boldsymbol{f}_i^{\text{int}}$ in \cref{eq:internal_force} in the next P2G step. 


\vspace{5pt}
\noindent\textbf{Disentanglement of Deformation Gradient.}\quad
In this paper, our key insight is to decouple the kinematic velocity from the deformation gradient update (\cref{eq:dg_update}). To do this, we decompose $\boldsymbol{v}_p$ into a kinematic velocity $\boldsymbol{v}_p^k  \in \mathbb{R}^{N \times 3}$, where $N$ denotes the number of particles.


As a first step, we compute a velocity grid $\boldsymbol{v}_i^k$ separated solely from the kinematic velocity.
Similar to the basic MPM (\cref{eq:p2g_mv}), we conduct P2G step to transfer $\boldsymbol{v}_p^k$ to a kinematic grid velocity field $\boldsymbol{v}_i^k$. By transferring $\boldsymbol{v}_p^k$ and its associated particle mass $m_p^k$ onto grid nodes, we compute their mass $m_i^k$ and kinematic momentum $m_i^k\boldsymbol{v}_i^k$:
\begin{equation}
    m_i^k = \sum_p w_{ip} \, m_p^k, \quad
    m_i^k\boldsymbol{v}_i^k =
    \sum_p w_{ip} \, m_p^k \boldsymbol{v}_p^k. \label{eq:mvk}
\end{equation}
By dividing the $m_i^k\boldsymbol{v}_i^k$ into $m_i^k$, we obtain $\boldsymbol{v}_i^k$, and then repeat the same procedures in \cref{eq:dg_update} as below:
\begin{align}
\nabla \boldsymbol{v}_p^k &= \sum_i \boldsymbol{v}_i^k \, (\nabla w_{ip})^\top, \quad
\boldsymbol{F}_p^k \leftarrow \left( \mathbf{I} + \Delta t \, \nabla \boldsymbol{v}_p^k \right) \boldsymbol{F}_p^k, \label{eq:update-Fpk}
\end{align}

Finally, we extract the kinematic deformation gradient $\boldsymbol{F}_p^k$ from the total deformation gradient $\boldsymbol{F}_{p}$ as a first-order approximation:
\begin{equation}
\boldsymbol{F}_p \leftarrow \boldsymbol{F}_p \, ({\boldsymbol{F}_p^k})^{-1}.  \label{eq:update-Fp}
\end{equation}
By applying the above \cref{eq:update-Fp} into \cref{eq:kirchhoff_stress}, we achieve the physically consistent stress responses, even with the conventional MPM framework.
This update eliminates the unnecessary stress accumulation and contributes to stable, real-time physical avatar representation, while also allowing physical interactions.

\subsection{Skeleton-based Pose Regression}
\label{sec:method2}
In the physically interactive simulation, external forces are exerted onto an avatar, which yields its non-rigid deformations, such as shapes and poses.
In Sec.~\ref{sec:method1}, we can apply the kinematic velocity to animate the avatar as intended.

Next, we describe how to compute the kinematic velocity.
Obviously, it can be obtained from a difference between adjacent pose sequences.
Unfortunately, deformations make it hard to track poses over time because the physical interactions change the current pose so that it differs from the intended pose.

Our approach is thus to efficiently compute the kinematic velocity from a difference between adjacent pose sequences in real-time, suitable for MPM-based frameworks: (1) embedding a set of skeleton particles into avatar's surface to track its poses; (2) calculating avatar's kinematic velocity from the tracked poses.
The embedded skeleton inherently represents joint structures, which ensures consistency between the human avatar surface and its poses.


\vspace{5pt}
\noindent\textbf{Skeletal Pose Extraction.}\quad
We introduce a direct approach that embeds a set of skeleton particles into the avatar's surface, where each joint is represented by a group of bone particles.
When external forces act on the avatar, these forces are transferred from the surrounding surface to the corresponding bones, causing their particles' positions to change accordingly.
%
By tracking the motion of these particles, we compute the pose of each bone, including its rotation $\mathbf{R}$ and translation $\mathbf{t}$.
This representation ensures stable pose consistency and smooth motion propagation across adjacent joints.
The computation is designed in a closed-form manner and solved via a simple least-squares optimization.
%
%

To compute the transformation matrix for each bone, we apply the Kabsch algorithm $K(\cdot)$~\cite{kabsch1976solution} to its corresponding bone particles as:
\begin{equation}
\mathbf{T}^{t}_\text{bone} = [\mathbf{R}^{t}_\text{bone}, \mathbf{t}^{t}_\text{bone}] = K\left(\boldsymbol{x}^{\text{cano}}_{\text{bone}}, \boldsymbol{x}^{t}_{\text{bone}}\right),
\label{eq:kabsch}
\end{equation}
where $\mathbf{T}^{t}_\text{bone} \in \mathbb{R}^{J \times 4 \times 4}$ is a set of per-joint transformation matrices at frame $t$. Here, $J$ is the number of joints.
It estimates the rotation $\mathbf{R}^{t}_\text{bone}$ and the translation $\mathbf{t}^{t}_\text{bone}$ to align the canonical bone particle positions $\boldsymbol{x}^{\text{cano}}_{\text{bone}}$ with their current positions $\boldsymbol{x}^{t}_{\text{bone}}$.

\vspace{5pt}
\noindent\textbf{Kinematic Velocity Calculation.}\quad
For subsequent frames, the avatar is animated according to the input pose transformation sequence $\hat{\mathbf{T}} \in \mathbb{R}^{T \times J \times 4 \times 4}$, where $T$ is the number of frames. The kinematic velocity of each particle, $\mathbf{v}^k_p$, is then computed from the positional differences between consecutive posed avatars.

For the stable kinematic velocity and the non-rigid physical interaction, we update the current pose using only incremental transformations in the sequence. This formulation maintains consistent and stable kinematic velocities even under pose deformations induced by external forces.
We first compute a relative (incremental) transformation from a current pose $\hat{\mathbf{T}}^{t}$ and its next pose $\hat{\mathbf{T}}^{t+1}$ as follows:
\begin{equation}
    \hat{\mathbf{T}}^{t \rightarrow t+1} = (\hat{\mathbf{T}}^{t})^{-1} \hat{\mathbf{T}}^{t+1}.
\end{equation}
We then update it as below:
\begin{equation}
    \mathbf{T}^{t+1}_\text{bone} = \hat{\mathbf{T}}^{t \rightarrow t+1} \mathbf{T}^{t}_\text{bone}.
    \label{eq:pose_next}
\end{equation}
Using \cref{eq:pose_next}, we obtain the consecutive avatar positions via Linear Blend Skinning (LBS) as follows:
\begin{equation}
    \boldsymbol{x}^t \mspace{-2mu} = \mspace{-2mu} \mathrm{LBS}(\mathbf{T}^{t}_{\text{bone}},\, \! \boldsymbol{x}^{\text{cano}}), ~
    \boldsymbol{x}^{t+1} \mspace{-2mu} = \mspace{-2mu} \mathrm{LBS}(\mathbf{T}^{t+1}_{\text{bone}},\, \! \boldsymbol{x}^{\text{cano}}).
    \label{eq:LBS}
\end{equation}

The kinematic velocity $\boldsymbol{v}^k_p$ for each particle is then determined from the positional difference as follows:
\begin{equation}
    \boldsymbol{v}^k_p = \frac{\boldsymbol{x}^{t+1} - \boldsymbol{x}^{t}}{\Delta t}. \label{eq:avatar_vk}
\end{equation}
Finally, we plug \cref{eq:avatar_vk} into \cref{eq:mvk} for general use.

\section{Experimental Results}
We demonstrate the superiority of the proposed PIAvatar over the basic MPM framework and its versatility through a variety of physically aware scenarios such as human-human and human-object interactions.

\subsection{Experiment Details}

We evaluate our method using two types of avatars:
(1) a clothed Gaussian-based avatar using Animatable Gaussians (AG)~\cite{li2024animatable} trained on 7 ActorsHQ~\cite{isik2023humanrf} models, and
(2) a parametric mesh-based avatar represented with SMPL-X~\cite{pavlakos2019expressive}. Each avatar consists of 300,000 particles on average for AG, and 10,475 particles for SMPL-X.
For animating avatars, we choose pose sequences from AMASS dataset~\cite{mahmood2019amass}.
To evaluate physically aware interactions, we use public 3D assets from BlenderNeRF~\cite{maximeraafat_BlenderNeRF} and Sketchfab\footnote{\url{https://sketchfab.com}}.
We embed kinematic skeletons inside avatars via OSSO~\cite{keller2022osso}, which provides joint hierarchy, pose alignment and skinning weights.
For computational efficiency, we downsample the 74,496 particles constituting the OSSO skeleton by 10× using voxel grid downsampling, thereby reducing both simulation time and the cost of the Kabsch computation.
PIAvatar is built upon PhysGaussian~\cite{xie2024physgaussian}, which is a baseline of our work to utilize the conventional MPM framework and support visualization of both meshes and 3D Gaussians. Our method runs 100 steps to render each frame. We adopt a $200^3$ background grid unless otherwise stated. 
All simulations are executed on an AMD Ryzen 9 7950X3D 16-Core processor with a single NVIDIA RTX 4090 GPU and 3DGRUT~\cite{loccoz20243dgrt, wu20253dgut} is used for visualization.

\begin{table}[t]
\caption{Performance comparison across simulation steps between a basic MPM and Ours. MSE/RMSE ($m$) and Acc@0.01 (ratio within 0.01$m$) are reported.}
\centering
\renewcommand{\arraystretch}{1.25}
\setlength{\tabcolsep}{4pt}
\scriptsize
{%
\begin{tabular}{cl|cccc|cccc}
\noalign{\hrule height 1pt}
\hline
\multirow{2}{*}{\textbf{Method}\vspace{-5pt}} & \multirow{2}{*}{\textbf{Metric}\vspace{-5pt}} & \multicolumn{4}{c|}{\textbf{Animatable Gaussians}} & \multicolumn{4}{c}{\textbf{SMPL-X}} \\
\cline{3-10}
&& \textbf{100} & \textbf{200} & \textbf{300} & \textbf{400}
&  \textbf{100} & \textbf{200} & \textbf{300} & \textbf{400} \\

\noalign{\hrule height 1pt}
\multirow{3}{*}{\tworow{Baseline}{MPM}} 
&MSE$\;\downarrow$       & 0.079 & 0.198 & 0.275 & 0.332 & 0.089 & 0.138 & 0.206 & 0.248 \\
&RMSE$\;\downarrow$      & 0.100 & 0.232 & 0.317 & 0.380 & 0.120 & 0.165 & 0.242 & 0.291 \\
&Acc@0.01$\;\uparrow$         & 0.097 & 0.041 & 0.020 & 0.020 & 0.358 & 0.181 & 0.086 & 0.036 \\
\hline
\multirow{3}{*}{Ours} 
&MSE$\;\downarrow$       & 0.019 & 0.027 & 0.036 & 0.046 & 0.013 & 0.022 & 0.031 & 0.039 \\
&RMSE$\;\downarrow$      & 0.023 & 0.034 & 0.043 & 0.056 & 0.016 & 0.025 & 0.034 & 0.044 \\
&Acc@0.01$\;\uparrow$         & 0.844 & 0.719 & 0.606 & 0.534 & 0.908 & 0.823 & 0.670 & 0.583 \\

\noalign{\hrule height 1pt}
\hline
\end{tabular}}
\vspace{3pt}
\vspace{-6pt}
\label{tab:comparison_baseline_vs_ours}
\end{table}

\subsection{Quantitative Evaluation Results} 

\noindent\textbf{Comparisons with Baseline MPM.}
We measure MSE, RMSE, and Accuracy at $\tau=0.01m$ between our PIAvatar and the basic MPM with respect to the target avatar positions calculated from the pose sequences. To show the robustness of our method, we conduct experiments under two conditions: (1) $20$-poses sequence for each of the $7$ AG actors; (2) SMPL-X with random body shapes (betas), two genders, and $20$ poses. 

As shown in Tab.~\ref{tab:comparison_baseline_vs_ours}, our method reliably reflects pose changes to the avatars regardless of the conditions. Since we leverage Kinematic Deformation Decoupling, the user-defined velocity drives the intended motion without errors from stress.
In contrast, the basic MPM suffers from the increasing deviation from the target position over time.

\begin{table}[t]
\caption{Kabsch-based pose alignment errors across simulation steps for Animatable Gaussians and SMPL-X. Lower is better.}
\centering
\renewcommand{\arraystretch}{1.25}
\scriptsize
\setlength{\tabcolsep}{6pt}
\resizebox{\linewidth}{!}
{%
\begin{tabular}{l|cccc|cccc}
\hline
\noalign{\hrule height 1pt}
\multirow{2}{*}{\textbf{Metric}\vspace{-5pt}} & \multicolumn{4}{c|}{\textbf{Animatable Gaussians}} & \multicolumn{4}{c}{\textbf{SMPL-X}} \\
\cline{2-9}
& \textbf{100} & \textbf{200} & \textbf{300} & \textbf{400}
& \textbf{100} & \textbf{200} & \textbf{300} & \textbf{400} \\
\noalign{\hrule height 1pt}

Root Rot.~($^\circ$)\;$\downarrow$       
& 0.137 & 0.144 & 0.162 & 0.154 
& 0.030 & 0.061 & 0.049 & 0.121 \\

Root Trl.~(m)$\;\downarrow$         
& 0.0006 & 0.0006 & 0.0006 & 0.0007 
& 0.0002 & 0.0004 & 0.0003 & 0.0010 \\

Rel.~Rot.~($^\circ$)\;$\downarrow$       
& 0.632 & 0.616 & 0.663 & 0.642 
& 0.097 & 0.205 & 0.202 & 0.320  \\

Rel.~Trl.~(m)\;$\downarrow$         
& 0.0041 & 0.0038 & 0.0044 & 0.0042
& 0.0005 & 0.0011 & 0.0011 & 0.0021 \\

Distance Error.~(m)\;$\downarrow$         
& 0.0057 & 0.0052 & 0.0056 & 0.0056
& 0.0022 & 0.0025 & 0.0034 & 0.0032 \\

\noalign{\hrule height 1pt}
\hline
\end{tabular}}
\vspace{2pt}
\vspace{-3pt}
\label{tab:kabsch_framewise}
\end{table}

\begin{figure*}[t]
    \centering
    \includegraphics[clip=true, width=1\linewidth]{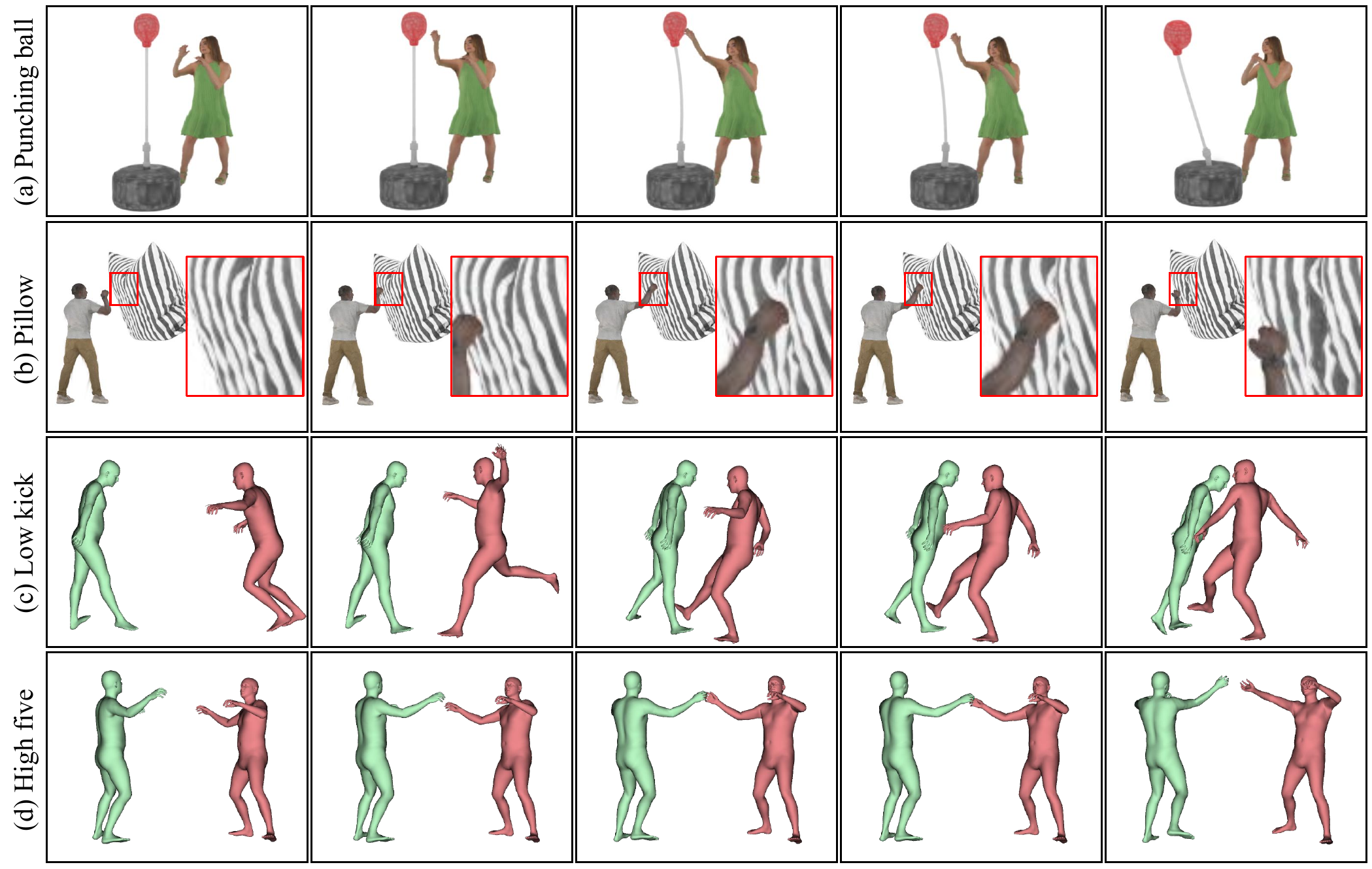}
    \vspace{-5mm}
    \caption{\textbf{Various Interactions.} (a, b) Non-rigid deformations arising from physical interactions with objects. (c, d) The bidirectional interaction between avatars and its mutual pose changes.}
    \vspace{-1pt}
    \label{fig:interactions}
\end{figure*}

\vspace{4pt}
\noindent\textbf{Pose Tracking.}
We evaluate the accuracy of continuous pose tracking using our skeletal model over the ground-truth pose. Specifically, we test 20 non-overlapping random poses for each of the seven AG models, and simulate a sequence of 200 random poses for the SMPL-X model. 

Tab.~\ref{tab:kabsch_framewise} describes the rotation and translation errors of both the root and relative joints. We compute velocities directly from the target-frame pose following the Half-Physics~\cite{siyao2025half} protocol, rather than relying on intermediate pose updates.
As shown in the results, our gradient decoupling strategy produces only negligible differences, at the level of floating-point precision. This verifies that the proposed formulation preserves skeletal pose consistency during continuous simulation.

\begin{table}[t]
\caption{Runtime comparison for single- and four-avatar interaction scenarios using SMPL-X and Animatable Gaussians. All times are measured in milliseconds (ms).}
\centering
\renewcommand{\arraystretch}{1.22}
\scriptsize
\setlength{\tabcolsep}{4pt}
{%
\begin{tabular}{l|cc|cc}
\hline
\noalign{\hrule height 1pt}
\textbf{Component} 
& \textbf{Single SMPL-X} 
& \textbf{Four SMPL-X} 
& \textbf{Single AG~\cite{li2024animatable}} 
& \textbf{Four AG} \\
\noalign{\hrule height 1pt}
Basic MPM                 & 0.170 & 0.311 & 2.619 & 11.015 \\
Velocity Computation      & 0.149 & 0.542 & 0.976 &  3.661 \\

\hline
+ Grid Initialization     & 0.151 & 0.518 & 0.145 & 0.472 \\
+ P2G                     & 0.0068 & 0.0153 & 0.0076 & 0.0043 \\
+ G2P                     & 0.0068 & 0.0170 & 0.0085 & 0.0748 \\
+ Kabsch algorithm        & 0.0085 & 0.0324 & 0.0082 & 0.0319 \\

\hline
\textbf{Total time (ms)}  & \textbf{0.492} & \textbf{1.436} & \textbf{3.764} & \textbf{15.259} \\
\noalign{\hrule height 1pt}
\hline
\end{tabular}}
\vspace{5pt}
\label{tab:runtime_multi}
\end{table}

\vspace{4pt}
\noindent\textbf{Time and Space Complexity.}
The detailed runtime is described in Tab.~\ref{tab:runtime_multi}. The experiments are conducted using single- and four-avatar scenarios in SMPL-X and AG. 
The velocity generation and the Kabsch algorithm are executed once per frame, so we divide them by the number of simulation steps per frame.
The overall overhead remains modest compared to the basic MPM, as the increased time cost in the grid initialization step is inevitable. 
This can be further improved with known techniques, such as fixing the number of additional grids and sharing them across avatars via an atomic CAS-based allocation strategy~\cite{herlihy1991wait}.
The additional memory overhead comes from an extra $200^3$ grid allocation per avatar to store kinetic momentum and their mass, which needs 122\,MB for both types of avatar. This is relatively small compared to the basic MPM memory usage of approximately 3.8\,GB for SMPL-X and 5.4\,GB for AG.


\subsection{Qualitative Evaluation Results}
Our physics-based avatar simulation framework enables diverse qualitative evaluations, showcasing interaction outcomes with a variety of physical capabilities.

\noindent\textbf{Human-Object Interactions.}
As shown in Fig.~\ref{fig:interactions}(a, b), the momentum naturally impacts both the avatars and the objects, leading to the temporary non-rigid deformations of the avatar or displacements of the objects through the transmitted forces.
These results validate that the user-defined kinematic velocity enables physically plausible interactions with the objects.

\noindent\textbf{Human-Human Interactions.}
In Fig.~\ref{fig:interactions}(c, d), the external forces generated by one avatar can be directly transmitted to another, altering its pose.
This enables physics-based interactions between avatars, including striking motions to lose balance or to be pushed away.

\noindent\textbf{Non-rigid Simulation.}
\begin{figure}[t]
    \centering
    \includegraphics[clip=true, width=1.0\linewidth]{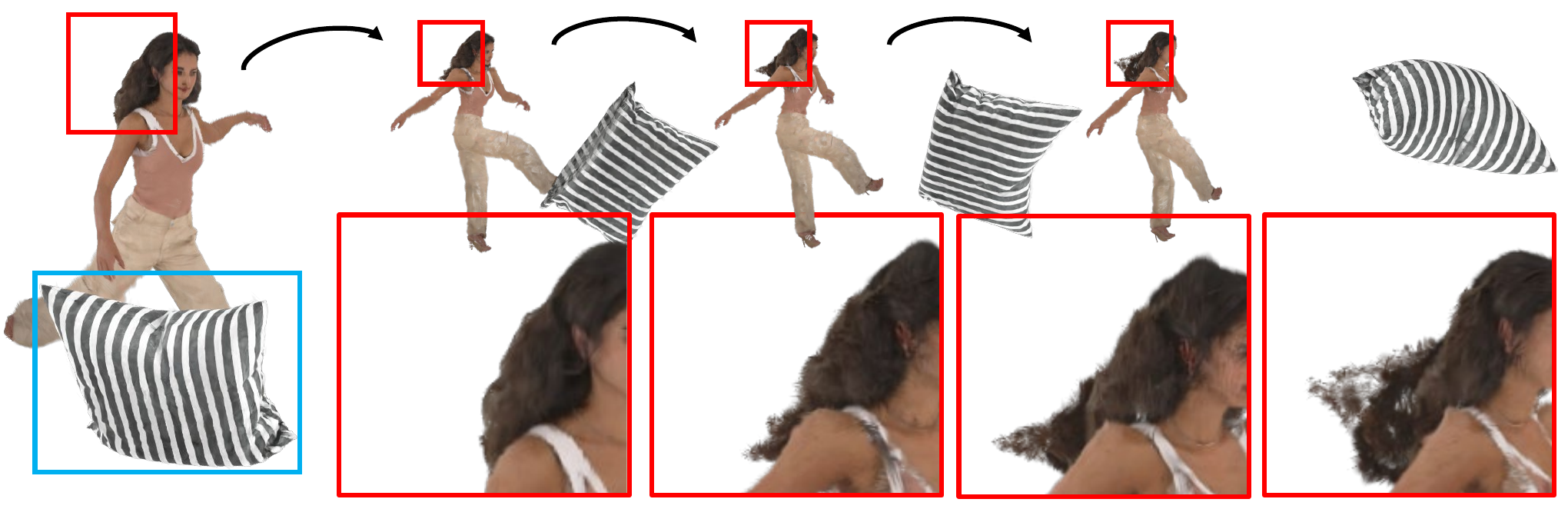}
    \caption{\textbf{Non-rigid deformation.} Our method can generate non-rigid deformations, such as naturally fluttering hair, that are not achievable with conventional avatars without explicit modeling.}
    \label{fig:Non-rigid}
\end{figure}
Fig.~\ref{fig:Non-rigid} shows that the momentum transmitted from a pillow has an impact on the avatar's body and hair. These results suggest that our physics-based simulation can reproduce such secondary effects, without explicitly modeling fine-scale deformations or using any additional hair simulation model.
Similarly, in Fig.~\ref{fig:interactions}(a, b), we have already displayed the physical momentum on non-rigid subjects.

\begin{figure*}[t]
\centering
\includegraphics[width=1.0\linewidth]{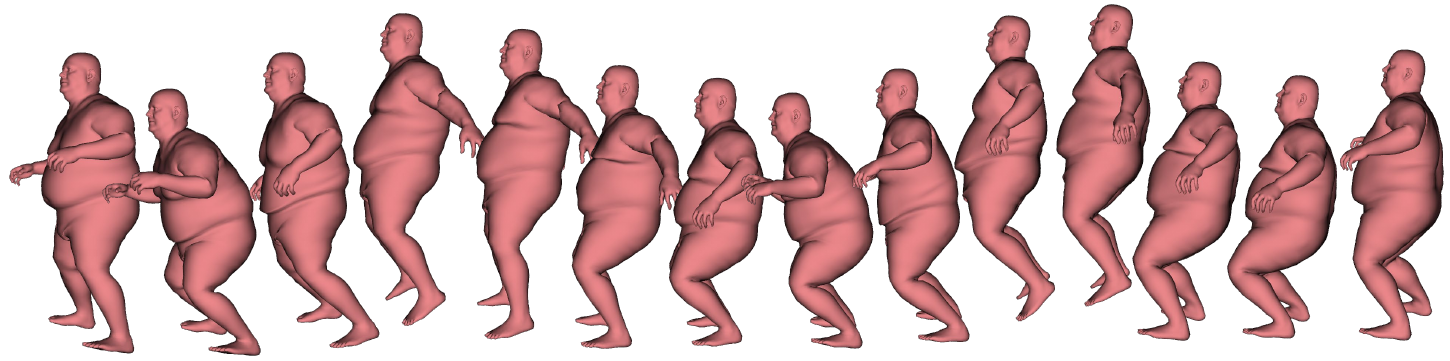}
\vspace{-15pt}
\caption{\textbf{Soft-tissue deformable avatar.}
Our simulator produces non-rigid effects, such as the natural jiggling of belly fat.
Unlike conventional LBS-based avatars, where body motion does not influence surface deformation, our physical simulation system exhibits natural inertial wobbling during jumping and landing.
}
\vspace{-10pt}
\label{fig:nonrigid}
\end{figure*}

\noindent\textbf{Soft-Tissue Deformation.} 
Fig.~\ref{fig:nonrigid} further demonstrates that our formulation reproduces soft-tissue deformation as a byproduct.
Without introducing any additional soft-body models, belly jiggling emerges naturally from particle interactions.
In practice, we slightly attenuate the kinematic velocity applied to particles in the belly region, allowing stresses to develop through neighboring interactions within the same unified framework.


\noindent\textbf{Deformation Gradient Visualization.}
\begin{figure}[t]
    \centering
    \includegraphics[clip=true, width=\linewidth]{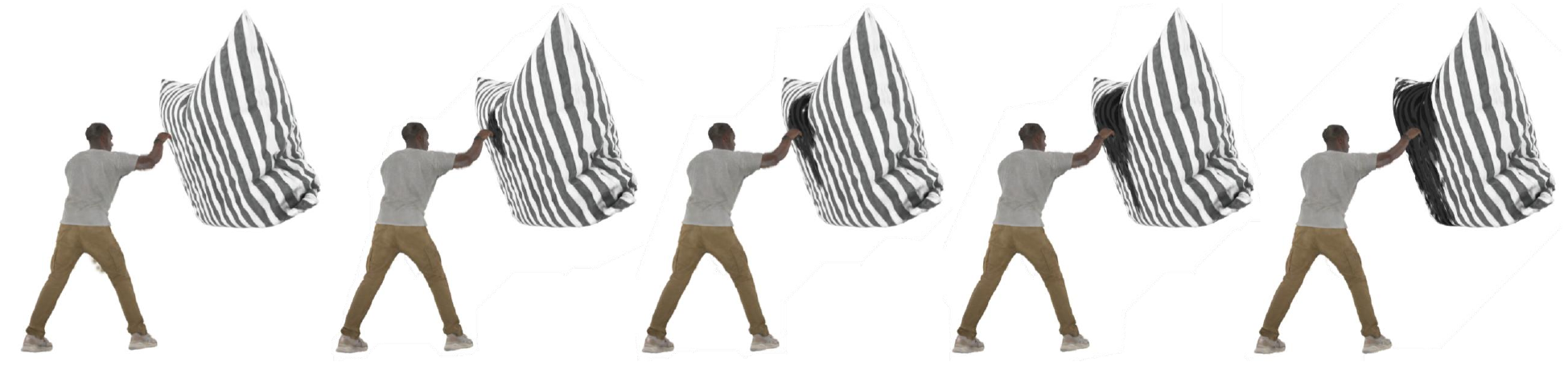}
    \vspace{-15pt}
    \caption{\textbf{Deformation gradient visualization.} The simulation visualizes how forces are transmitted through changes in the deformation gradient $\boldsymbol{F}_p$.}
    \label{fig:F_visual}
    \vspace{-7pt}
\end{figure}
We visualize how a deformation gradient of a non-rigid object changes in Fig.~\ref{fig:F_visual}.
During the interaction, starting from the hitting point of the object, the deformation gradient progressively propagates outward to the surrounding area.

\noindent\textbf{Heterogeneous Material Parameters.}
\begin{figure}[!t]
    \centering
    \includegraphics[clip=true, width=\linewidth]{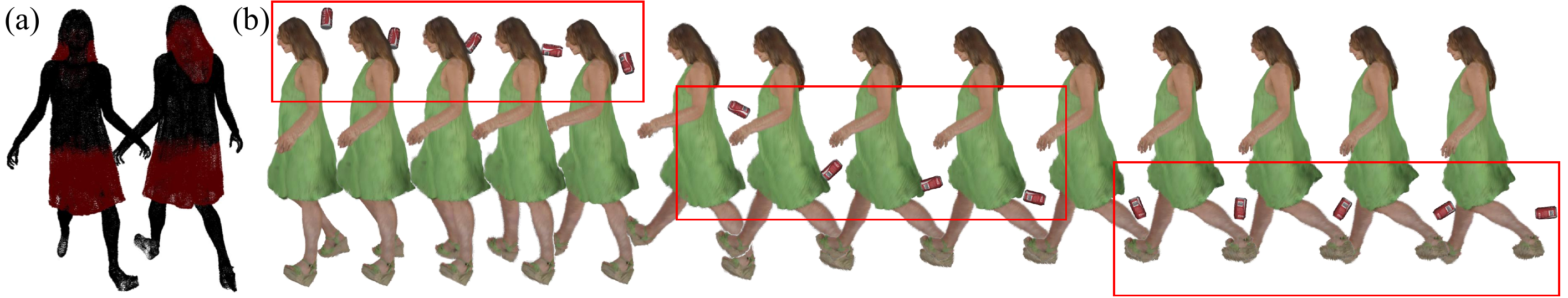}
    \vspace{-19pt}
    \caption{
    \textbf{Heterogeneous material assignment.}
    (a) The body (Neo-Hookean, $E{=}10^5$) and compliant regions (corotated, $E{=}10^2$–$10^3$) exhibit distinct deformation magnitudes under motion.
    (b) Despite regional variation in stiffness, global skeletal articulation remains stable and consistent with the input pose sequence.
    }
    \vspace{-5pt}
    \label{fig:heterogeneous}
\end{figure}
We lastly evaluate the versatility of our PIAvatar through an experiment on a heterogeneous material assignment. The body is modeled using a Neo-Hookean hyperelastic model ($E{=}10^5$), while garments and hair are assigned more compliant corotated linear elastic models ($E{=}10^2$ and $10^3$), introducing relatively different stiffness according to the parts of the avatar (Fig.~\ref{fig:heterogeneous}(a)). These values are calibrated as effective engineering parameters to ensure numerical stability within the MPM discretization. As shown in Fig.~\ref{fig:heterogeneous}(b), we observe different collision responses when the fallen coke can hits the hair, garment, and the ankle.

\begin{figure}[!t]
    \centering
    \includegraphics[clip=true, width=\linewidth]{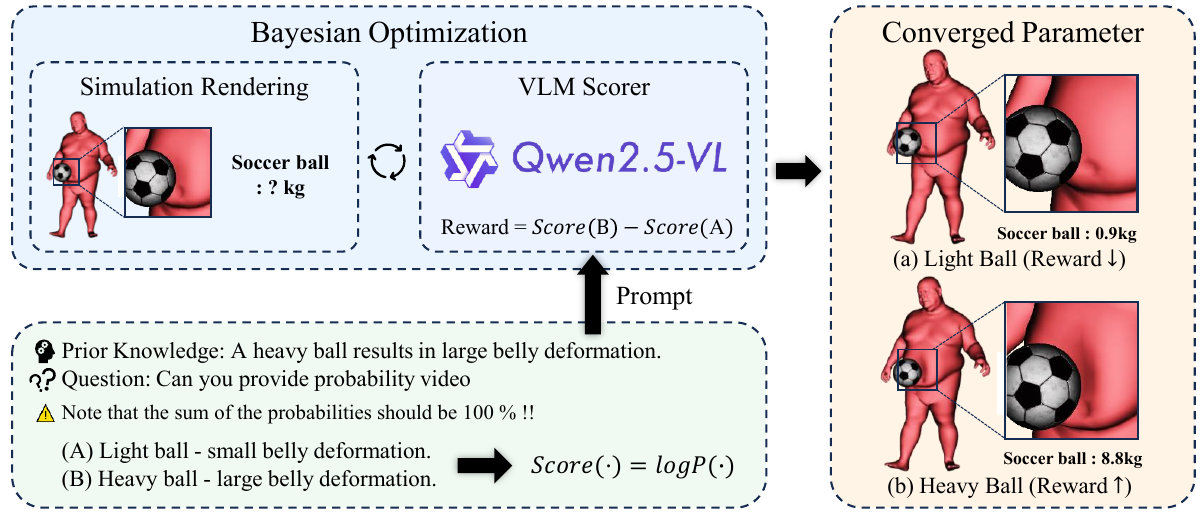}
    \vspace{-17pt}
    \caption{
    \textbf{Automatic physical parameter optimization.} The difference between the log probabilities assigned by the VLM to the binary answer, such as \emph{light} vs \emph{heavy}, is used as the BO reward for automatic parameter settings.}
    \vspace{-3pt}
    \label{fig:BO}
\end{figure}

\subsection{VLM-Driven Physical Parameter Optimization}
\label{sec:bo_vlm}

We introduce a potential way to automatically find physical parameters using Vision-Language Models~(VLM).
As shown in Fig.~\ref{fig:BO}, we design an additional experiment that optimizes the ball's density $d$ or the belly's stiffness $E$ with both Bayesian optimization (BO)~\cite{frazier2018bayesian} and Qwen2.5-VL~\cite{Qwen2.5-VL} as the VLM reward model.


Since VLMs lack direct intuition for abstract physical quantities, we prepare prompts for an observable cues, e.g., larger belly deformation for a heavier ball or smaller deformation for a stiffer belly as a prior knowledge.
As illustrated in Fig.~\ref{fig:BO}, at each BO iteration, the VLM is asked to choose between two contrasting prompts (\eg, \emph{heavy} vs.\ \emph{light}, or \emph{stiff} vs.\ \emph{soft}) for the rendered simulation video. The difference between the log probabilities of the two answers is then used as the BO reward.
We optimize physical parameters with BO, density $d$\footnote{Object's mass(kg) is computed from the multiplication of its density $d$(kg/m$^3$) and volume(m$^3$).} and Young's modulus $E$: maximizing the reward for \emph{heavy}/\emph{stiff} behavior leads to the larger physical parameters, whereas minimizing the same reward leads to smaller ones corresponding to \emph{light}/\emph{soft} behavior.
In Tab.~\ref{tab:bo_vlm_rebuttal}, the proposed optimization framework consistently converges, validating that the VLM-based reward design successfully distinguishes heaviness/lightness as positive/negative scores, respectively.

We evaluate the proposed BO with the VLM framework through a user study with $50$ participants in MTurk. 
We manually generate videos with six different attributes, and the participants are asked to choose most plausible videos under the prompts regarding to ball weight or belly softness.
As reported in Tab.~\ref{tab:user_study_votes}, the most-chosen videos closely match the BO-converged values in the red/blue cells of Tab.~\ref{tab:bo_vlm_rebuttal}, implying that the VLM-driven proxy aligns with human perception.

\begin{table}[t]
\centering
\renewcommand{\arraystretch}{1.22}
\scriptsize
\setlength{\tabcolsep}{1pt}
\begin{minipage}{0.54\linewidth}
\vspace{-8pt}
\captionof{table}{BO with VLM scorer converges to opposite parameter per the text target.}
\vspace{-8pt}
\centering
\resizebox{\linewidth}{!}{%
\begin{tabular}{l|cc|cc}
\hline
\noalign{\hrule height 1pt}
\textbf{Axis}
& \textbf{Converged} & \textbf{Reward}
& \textbf{Converged} & \textbf{Reward} \\
\noalign{\hrule height 1pt}
Mass (kg)
& \cellcolor{red!25}$\mathbf{8.8}$ \emph{(heavy)} & $+3.53$
& \cellcolor{blue!25}$\mathbf{0.9}$ \emph{(light)} & $-2.90$ \\
$E$ (Pa)
& \cellcolor{red!25}$\mathbf{3.38{\times}10^{6}}$ \emph{(stiff)} & $+1.17$
& \cellcolor{blue!25}$\mathbf{3.26{\times}10^{2}}$ \emph{(soft)} & $-0.81$ \\
\noalign{\hrule height 1pt}
\hline
\end{tabular}}
\label{tab:bo_vlm_rebuttal}
\end{minipage}
\hfill
\begin{minipage}{0.45\linewidth}
\vspace{-4pt}
\captionof{table}{User study. Modal selections (\textcolor{red!55}{red}/\textcolor{blue!55}{blue}) match the BO in Tab.~\ref{tab:bo_vlm_rebuttal}.}
\vspace{-9.5pt}
\centering
\resizebox{\linewidth}{!}{%
\begin{tabular}{l|cccccc}
\hline
\noalign{\hrule height 1pt}
Mass (kg) & 0.5 & \cellcolor{blue!25}\textbf{1.0} & 1.7 & 3.7 & \cellcolor{red!25}\textbf{8.2} & 10.0 \\
\emph{heavy\,/\,light} & 1\,/\,12 & \cellcolor{blue!25}1\,/\,\textbf{20} & 4\,/\,7 & 13\,/\,8 & \cellcolor{red!25}\textbf{16}\,/\,1 & 15\,/\,2 \\
\cline{1-7}
$E$ (Pa) & $10^{2}$ & \cellcolor{blue!25}{$\mathbf{4.2{\cdot}10^{2}}$} & $3.7{\cdot}10^{3}$ & $1.3{\cdot}10^{5}$ & \cellcolor{red!25}{$\mathbf{2.4{\cdot}10^{6}}$} & $10^{7}$ \\
\emph{stiff\,/\,soft} & 4\,/\,6 & \cellcolor{blue!25}4\,/\,\textbf{16} & 5\,/\,12 & 4\,/\,9 & \cellcolor{red!25}\textbf{23}\,/\,2 & 10\,/\,5 \\
\noalign{\hrule height 1pt}
\hline
\end{tabular}}
\label{tab:user_study_votes}
\end{minipage}
\vspace{-15pt}
\end{table}

\subsection{Loose Cloth Simulation} 
\label{sec:exp_loose_cloth}

\begin{figure}[t!]
    \centering
    \includegraphics[clip=true, width=\linewidth]{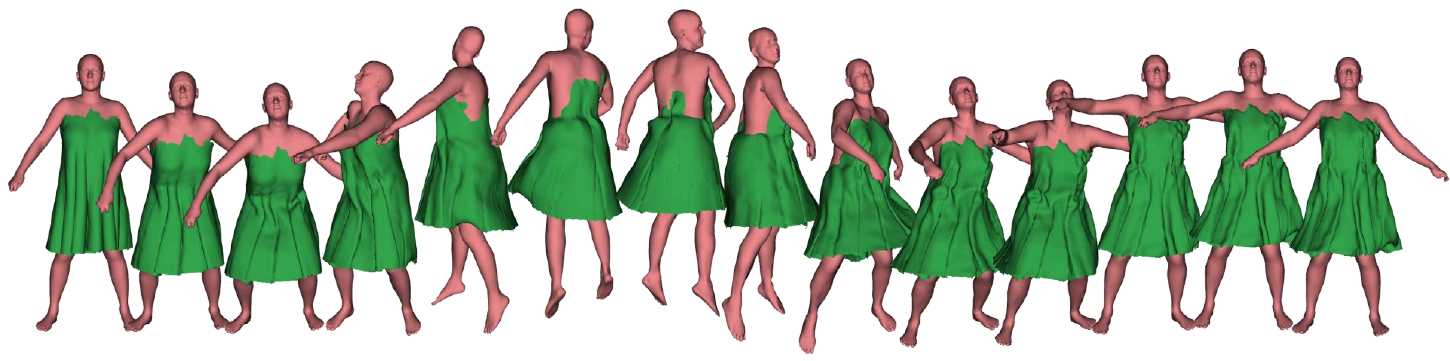}
    \vspace{-23pt}
    \caption{
    \textbf{Loose-garment simulation.} On \emph{spin} and \emph{jump} from fast AMASS motion sequences, our coupled cloth solver produces visible flutter and asymmetric billowing of the garment driven by rapid body articulation.}    
    \vspace{-16pt}
    \label{fig:loose_cloth_sim}
\end{figure}

Existing LBS-based avatars deform garments using skinning weights inherited from the underlying body, so loose clothing rigidly follows the skin and cannot exhibit secondary dynamics such as swing, lag or inertia.
Leveraging the recent MPMAvatar cloth solver~\cite{lee2025mpmavatar} and the extensibility of our MPM-based simulator, we apply our framework to loose-garment scenarios by coupling a thin-shell cloth solver with our MPM body simulator.
For garments, we replace volumetric MPM with a separate thin-shell formulation that applies a QR decomposition to the per-element deformation gradient, decoupling in-plane membrane stretching from out-of-plane bending.
As shown in Fig.~\ref{fig:loose_cloth_sim}, we show the example of loose-garment simulation that the dress flutters naturally.

\vspace{-5pt}
\section{Conclusion}
We introduce PIAvatar, an MPM-based avatar simulation framework that achieves physically aware, non-rigid and deformable avatar interactions as well as their surrounding objects.
By (1) decoupling the deformation gradient and (2) embedding a skeletal structure for direct pose estimation, PIAvatar enables stress-free kinematic control and demonstrates physically consistent interaction behaviors across diverse scenarios.

\noindent\textbf{Limitation.}
To create desired physical interaction scenarios, the manual setup of avatar positions and pose sequence is required. The current pose datasets, such as AMASS~\cite{mahmood2019amass}, lack a torque control under external forces like gravity, and friction is not implemented, making posture maintenance difficult. 
Potential solutions include PID control or RL for the torque control, and integrating friction and object properties through large-scale physics engines like Genesis~\cite{Genesis}. 
In addition, rapid collisions between two objects can also cause them to stick before stress develops. 
This raises the need for a new learning framework for understanding scene configurations and object characteristics.

\vspace{5pt}
\footnotesize{
\noindent\textbf{Acknowledgement}
This work was supported by the National Research Foundation of Korea (NRF) grant (RS-2024-00338439), the Institute for Information $\&$ Communications Technology Planning $\&$ Evaluation (IITP) grant funded by the Korea government (MSIT) (No.RS-2025-25441838, Development of a human foundation model for human-centric universal artificial intelligence and training of personnel), and the Korea Institute of Science and Technology (KIST) Institutional Program (26K0030).
}

\newpage

\begin{center}
  {\Large\bfseries
   PIAvatar: Physically Interactive Avatars via\\
   Deformation Gradient Decoupling\par}
  \vspace{10pt}
  {\large\bfseries Supplementary Material\par}
\end{center}
\vspace{-10pt}

\begin{figure*}[ht!]
\centering
\includegraphics[width=1.0\linewidth]{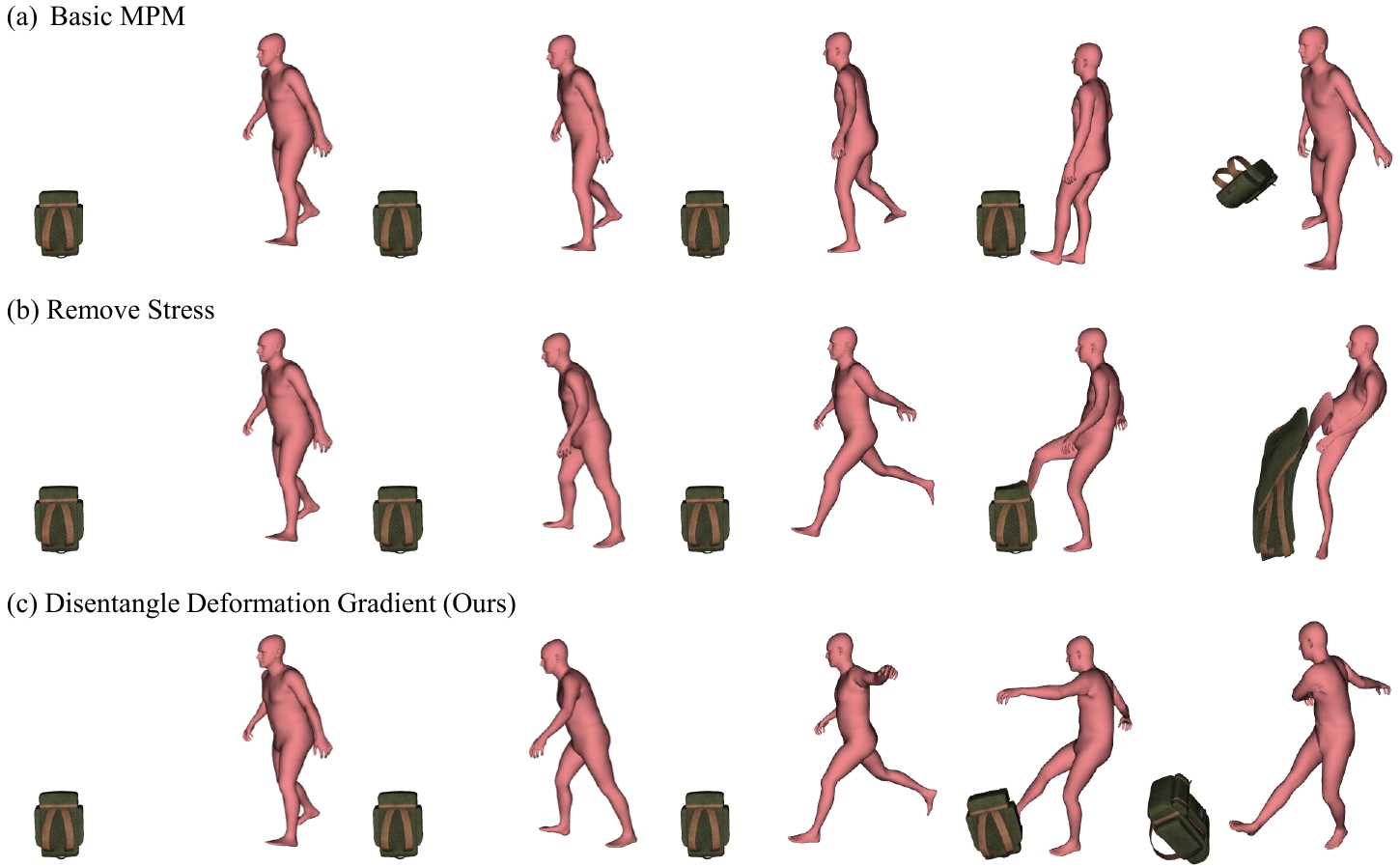}
\vspace{-15pt}
\caption{\textbf{Motivation for our disentangled deformation gradient.}
(a) Basic MPM fails to reach the target pose due to internal stress.
(b) Removing stress improves pose following but loses contact responses, causing merging with objects.
(c) Our method removes kinematic-induced stress while keeping external contact stress, enabling both pose tracking and physical interactions.}
\vspace{-5mm}
\label{fig:motivation}
\end{figure*}

\vspace{-20pt}
\section{Implementation Details}
Results presented in the main paper and this supplementary material are also provided as video demonstrations on the project page.

\subsection{Motivation and Design Rationale} 
\noindent\textbf{Pose-based Control.} To explain why we control the avatar using poses, we note that muscle-activation based control exists as an alternative, but it comes with limitations.
One key issue is that the activation → torque → pose mapping is highly nonlinear, which makes the inverse mapping (pose → activation) complex and prevents reliable control toward a desired pose.
In addition, some desired poses cannot be reached, for example as reported in~\cite{lee2018dexterous}, where raising an arm above the head cannot be reliably achieved.
For flexibility and controllability, we therefore adopt pose-based control.

\noindent\textbf{Decoupling the Deformation Gradient.}
Directly applying pose-driven velocities within a physical simulator introduces its own challenges, motivating the need to decouple the deformation gradient.
Fig.~\ref{fig:motivation} illustrates the motivation behind our disentangled deformation gradient design.
Fig.~\ref{fig:motivation}(a) shows that, in a basic MPM setup, directly applying velocity generates internal stress that prevents the avatar from reaching the target pose and leaves it with almost no visible pose deformation.
As shown in Fig.~\ref{fig:motivation}(b), removing all stress enables the avatar to partially follow the intended motion, but it cannot push against objects and instead merges with them.
To retain physical interactions, we preserve stress arising from external contacts while excluding stress induced by kinematic velocity, as demonstrated in Fig.~\ref{fig:motivation}(c). Through this disentangled formulation, the avatar can also achieve the desired pose deformation.

\begin{algorithm}[t]
\caption{Compute Kinematic Velocity of Avatar}
\label{alg:human_velocity}
\begin{spacing}{1.2}
\begin{algorithmic}[1]
\newcommand{\AlgNote}[1]{\Statex \(\triangleright\) #1}
\algblockdefx{Block}{EndBlock}[1]{\textbf{#1}}
  
\Require current rotations $\mathbf{R}^{t}_{\text{bone}}$ and translations $\mathbf{t}^{t}_{\text{bone}},$ canonical joints $\mathbf{J}$,
\Statex \hspace*{2.75em} current and next pose dataset $\hat{\mathbf{T}}^{t}, \hat{\mathbf{T}}^{t+1},$ number of joints $J$,
\Statex \hspace*{2.75em}   joint parents $p$, linear blend skinning LBS,
\Statex \hspace*{2.75em} avatar canonical positions $\boldsymbol{x}^{\text{cano}}$, time step $\Delta t$


\Ensure
kinematic velocity $\boldsymbol{v}^{t}_{k}$

\State $\hat{\mathbf{R}}^{t}, \hat{\mathbf{t}}^{t} \gets \hat{\mathbf{T}}^{t}$  \Comment{\textbf{Current input sequence}}
\State $\hat{\mathbf{R}}^{t+1}, \hat{\mathbf{t}}^{t+1} \gets \hat{\mathbf{T}}^{t+1}$ \Comment{\textbf{Next input sequence}}
\State $\hat{\mathbf{R}}^{t \to t+1} \gets (\hat{\mathbf{R}}^{t})^{-1} \hat{\mathbf{R}}^{t+1} $ \Comment{\textbf{Relative rotation}}

\State $\mathbf{R}^{t+1}_\text{bone} \gets \hat{\mathbf{R}}^{t\to t+1} \mathbf{R}^{t}_\text{bone}$


\State $\mathbf{J}_{\text{rel}} \gets \mathbf{J}$ \Comment{\textbf{Initialize relative joints}}
\State $\mathbf{J}_{\text{rel}}[1{:}] \gets \mathbf{J}[1{:}] - \mathbf{J}\bigl[p[1{:}]\bigr]$ \Comment{\textbf{Relative joint offset}}
\State $\mathbf{t}^{t+1}_\text{bone}[0] \gets \mathbf{J}_{\text{rel}}[0]$

\For{$i = 1$ \textbf{to} $J$} \Comment{\textbf{Hierarchical update}}
\State $\mathbf{t}^{t+1}_\text{bone}[i] \gets \mathbf{t}^{t+1}_\text{bone}[p[i]] + \mathbf{R}^{t+1}_\text{bone}[p[i]] \mathbf{J}_{\text{rel}}[i]$\label{line:transl}
\EndFor

\State $\mathbf{t}^{t}_{\text{bone}} \gets \mathbf{t}^{t}_{\text{bone}} - \mathbf{R}^{t}_\text{bone} \mathbf{J}$ \Comment{\textbf{Global translation}}
\State $\mathbf{t}^{t}_\text{bone}[0] \gets \mathbf{t}^{t}_\text{bone}[0] - \mathbf{J}[0] + \mathbf{R}^{t}_\text{bone}[0] \mathbf{J}[0]$
\State $\mathbf{t}^{t+1}_\text{bone} \gets \mathbf{t}^{t+1}_\text{bone} - \mathbf{R}^{t+1}_\text{bone} \mathbf{J} + (\hat{\mathbf{t}}^{t+1} - \hat{\mathbf{t}}^{t} + \mathbf{t}^{t}_\text{bone})[0]\mathbf{1}$

\State $\mathbf{T}^{t+1}_\text{bone}\gets[\mathbf{R}^{t+1}_\text{bone}|\mathbf{t}^{t+1}_\text{bone}], \mathbf{T}^{t}_\text{bone}\gets[\mathbf{R}^{t}_\text{bone}|\mathbf{t}^{t}_\text{bone}]$
\State $\boldsymbol{x}^t \mspace{-2mu} = \mspace{-2mu} \mathrm{LBS}(\mathbf{T}^{t}_{\text{bone}},\, \! \boldsymbol{x}^{\text{cano}}), ~
    \boldsymbol{x}^{t+1} \mspace{-2mu} = \mspace{-2mu} \mathrm{LBS}(\mathbf{T}^{t+1}_{\text{bone}},\, \! \boldsymbol{x}^{\text{cano}})$
\State $\boldsymbol{v}^k = ({\boldsymbol{x}^{t+1} - \boldsymbol{x}^{t}})/{\Delta t}$
\State \textbf{return} $\boldsymbol{v}_{k}$






\end{algorithmic}
\end{spacing}
\end{algorithm}
\vspace{-15pt}


\vspace{6pt}
\subsection{Details of Kinematic Velocity Calculation} 
We describe the velocity computation introduced in Sec.~\ref{sec:method2} using Alg.~\ref{alg:human_velocity}.
Pose-sequence datasets such as AMASS~\cite{mahmood2019amass} provide per-frame joint transformation matrices (lines 1–2).
We compute the relative joint rotation $\hat{\mathbf{R}}^{t \to t+1}$ (line 3) and apply it to the current bone rotation $\mathbf{R}^{t}_\text{bone}$ to obtain $\mathbf{R}^{t+1}_\text{bone}$ (line 4).
Following the SMPL-X~\cite{SMPLX_2019_Pavlakos} convention, we then compute the distances between relative joints (lines 5–6).
Using this hierarchy, the translations of all joints in the next frame are computed recursively (lines 7–10), ensuring stable pose consistency by preserving inter-joint distances.
Next, we extract the global translation from the root joint translation $\mathbf{t}^{t}_\text{bone}[0]$ (lines 11–12) and add the translation increment provided by the pose-sequence dataset to all joints (line 13).
Finally, we construct the full transformation matrix for each joint (line 14) and compute the kinematic velocity $\boldsymbol{v}_k$ via linear blend skinning (lines 15–16).
The resulting transformation matrices remain fully compatible with the standard SMPL-X formulation.



\begin{algorithm}[t]
\caption{Kabsch Algorithm}
\label{alg:kabsch}
\begin{spacing}{1.1}
\begin{algorithmic}[1]

\Require point sets $\mathbf{P},\ \mathbf{Q} \in \mathbb{R}^{n \times 3}$
\Ensure rotation $\mathbf{R}_\text{bone} \in \mathbb{R}^{3 \times 3}$, translation $\mathbf{t}_\text{bone} \in \mathbb{R}^{3}$

\State $\mathbf{P}_{\text{mean}} \gets \frac{1}{n} \sum_{i=1}^{n} \mathbf{P}[i]$
\State $\mathbf{Q}_{\text{mean}} \gets \frac{1}{n} \sum_{i=1}^{n} \mathbf{Q}[i]$

\State $\mathbf{P}_{\text{center}} \gets \mathbf{P} - \mathbf{P}_{\text{mean}}$ \Comment{\textbf{Center the point sets}}
\State $\mathbf{Q}_{\text{center}} \gets \mathbf{Q} - \mathbf{Q}_{\text{mean}}$

\State $\mathbf{C} \gets \mathbf{Q}_{\text{center}}^\top \mathbf{P}_{\text{center}}$ 
\Comment{\textbf{Correlation matrix} $\mathbf{C}$}

\State $(\mathbf{U},\, \mathbf{S},\, \mathbf{V}^\top) \gets \mathrm{SVD}(\mathbf{C})$

\State $\mathbf{R}_\text{bone} \gets \mathbf{U}\, \mathbf{V}^\top$ 
\Comment{\textbf{Rotation} $\mathbf{R}_\text{bone}$}

\If{$\det(\mathbf{R}_\text{bone}) < 0$}
  \State $\mathbf{U}[:, -1] \gets -\mathbf{U}[:, -1]$ 
  \Comment{\textbf{Reflection correction}}
  \State $\mathbf{R}_\text{bone} \gets \mathbf{U}\, \mathbf{V}^\top$
\EndIf

\State $\mathbf{t}_\text{bone} \gets \mathbf{Q}_{\text{mean}} - \mathbf{R}_\text{bone}\, \mathbf{P}_{\text{mean}}$ 
\Comment{\textbf{Translation} $\mathbf{t}_\text{bone}$}

\State \textbf{return} $(\mathbf{R}_\text{bone},\, \mathbf{t}_\text{bone})$

\end{algorithmic}
\end{spacing}
\end{algorithm}
\vspace{-15pt}

\vspace{5pt}
\subsection{Details of Kabsch Algorithm} 

LBS-based avatars compute surface coordinates from each joint’s transformation matrix, making the estimation of joint rotations and translations extremely important.
In Sec.~\ref{sec:method2}, we embed a skeleton model inside the avatar for pose estimation, allowing each bone to represent a joint and enabling skeletal pose extraction to predict its rotation and translation.
To obtain fast and accurate rigid transformations $(\mathbf{R}^t_\text{bone}, \mathbf{t}^t_\text{bone})$ for each bone, we use the Kabsch algorithm~\cite{kabsch1976solution} . 
Similar rigid alignment methods, such as shape matching~\cite{muller2005meshless}, can also be used as alternatives.

In Alg.~\ref{alg:kabsch}, we describe the Kabsch algorithm used in Eq.~\ref{eq:avatar_vk} of Sec.~\ref{sec:method2}.
For convenience, given the canonical and posed bone points
$x^\text{cano}_\text{bone}$ and $x^{t}_\text{bone} \in \mathbb{R}^{n \times 3}$,
we denote them as $\mathbf{P}$ and $\mathbf{Q}$, respectively.
We apply this procedure to each bone to compute its rigid transformation $(\mathbf{R}^t_\text{bone}, \mathbf{t}^t_\text{bone})$, which provides the rotation and translation used for skeleton alignment in our kinematic model.

\section{Mitigating Simulation Artifacts}

\begin{figure*}[t]
\centering
\includegraphics[width=1.0\linewidth]{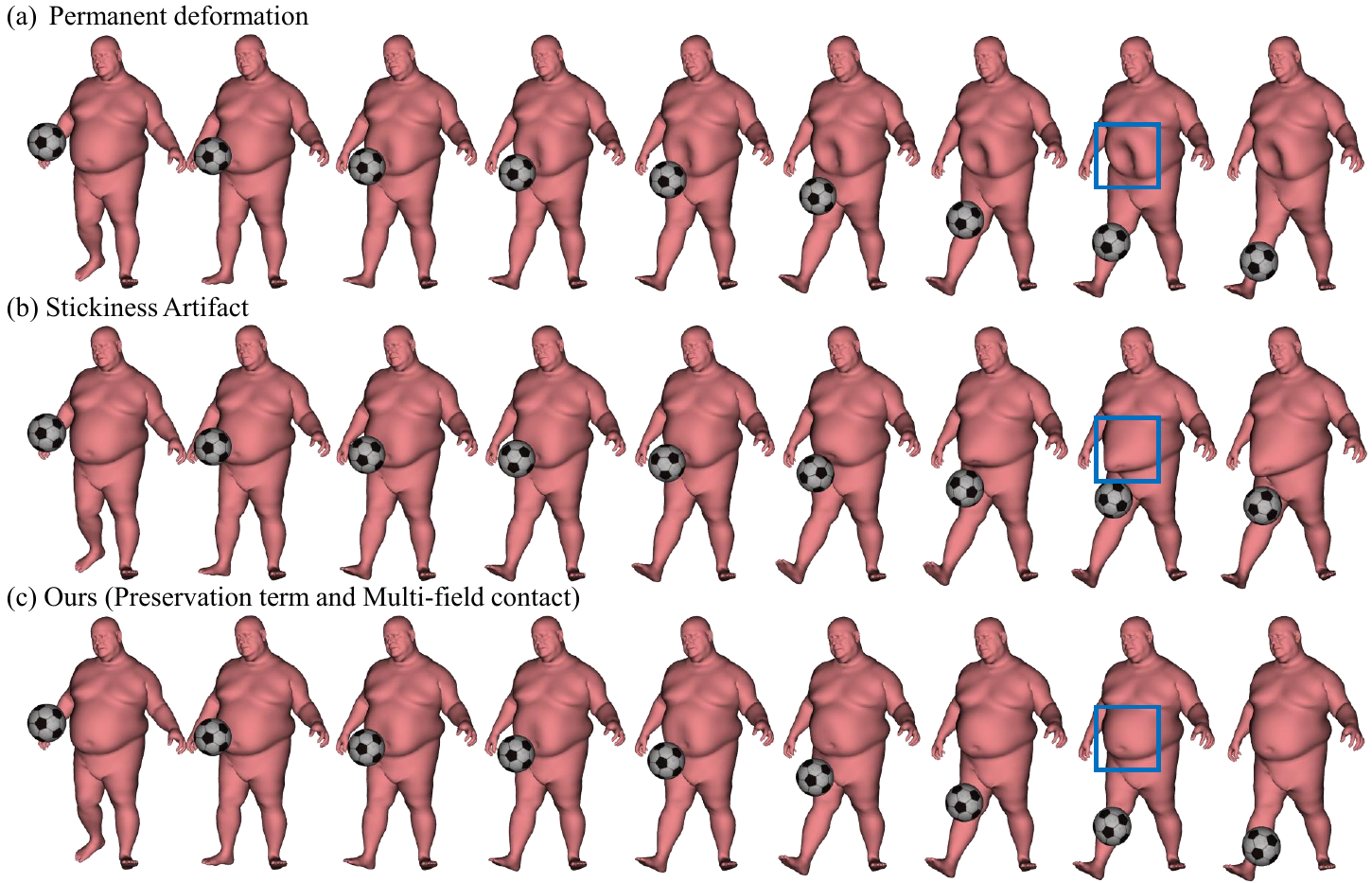}
\vspace{-20pt}
\caption{\textbf{Resolving Permanent deformation and Stickiness Artifacts}.
(a) Without the shape-preservation term, the belly retains a residual dent after the ball passes.
(b) With single-field MPM contact, the belly is dragged along with the ball and stretches instead of releasing.
(c) Combining the shape-preservation term with multi-field contact recovers the body shape and lets the ball detach cleanly.}
\vspace{-3mm}
\label{fig:artifact_alleviation}
\end{figure*}

\begin{figure*}[t]
\centering
\includegraphics[width=1.0\linewidth]{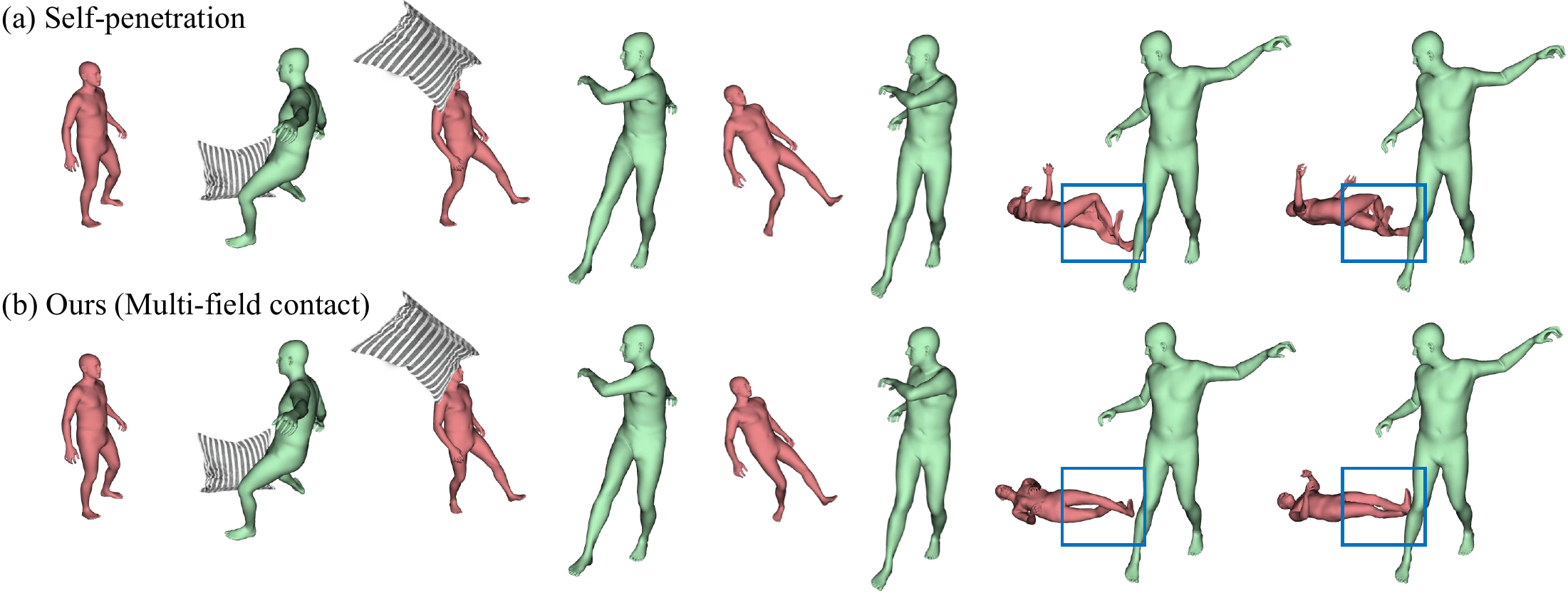}
\vspace{-10pt}
\caption{\textbf{Resolving Self-penetration Artifact}.
(a) With single-field MPM contact, the red subject's two legs fuse together once they touch and lose their individual shape.
(b) Our multi-field contact keeps the legs as distinct bodies on contact, preserving each leg's shape while interactions proceed naturally.}
\vspace{-3mm}
\label{fig:self_penetration}
\end{figure*}

\vspace{4pt}
\subsection{Mesh Topology Preservation.}
\label{supp:surface}
During MPM-based interactions, strong contact can deform the avatar surface and prevent it from recovering after contact. To avoid permanent deformation, we explicitly keep the avatar surface close to its posed shape. 
We preserve SMPL-X topology by adding our shape-preservation term to Eq.~\ref{eq:avatar_vk} inspired by meshless shape matching~\cite{muller2005meshless}:
%
\begin{equation}
    v_p^{k} \;=\; \underbrace{\frac{x_p^{\mathrm{LBS},t+1}-x_p^{\mathrm{LBS},t}}{\Delta t}}_{\text{kinematic LBS drive}}
    \;+\; \underbrace{\alpha\,\frac{x_p^{\mathrm{LBS},t+1}-x_p^{\mathrm{sim},t}}{\Delta t}}_{\text{shape-preservation}},
\label{eq:shape_pres}
\end{equation}
with $\alpha\in[0,1]$. 
When particles drift after contact, this term pulls them back toward the posed SMPL-X surface. As shown in Fig.~\ref{fig:artifact_alleviation}(a,~c), adding this term allows the avatar shape to recover after strong collisions.

\vspace{4pt}
\subsection{Multi-Field Contact for Stickiness and Self-Penetration.}
The stickiness artifacts in multi-subject contact arise from the same limitation of standard MPM. 
Also, ours approach relieves stress between body parts to enable natural movement, while allowing for self-insertion without explicit contact manipulation. 
In standard MPM, all subjects that contribute to the same grid node are merged into a single mass and momentum field and assigned one averaged grid velocity. 
Thus, the grid no longer distinguishes which subject contributed which momentum. 
As a result, they cannot properly limit their relative approach velocity, which can lead to interpenetration, and once their velocities are blended in same grid, they may also fail to separate, producing sticky contact.
Although interpenetration is often addressed using additional repulsion or penalty terms, we instead adopt a subject-wise momentum-preserving contact treatment. 

We address both artifacts using the multi-field contact formulation of Bardenhagen~et~al.~\cite{bardenhagen2000material}. 
Instead of storing a single mass and momentum at each grid node, we store subject-specific quantities $\{m_k,\mathbf{p}_k\}$ and compute a separate velocity for each subject,
$\mathbf{v}_k=\mathbf{p}_k/m_k$. 
For grid nodes influenced by multiple subjects, we first compute the center-of-mass velocity
$\mathbf{v}_\mathrm{cm}=\frac{\sum_k m_k\mathbf{v}_k}{\sum_k m_k}$.
We then estimate an outward normal for each subject from the negative gradient of its grid mass,
$\mathbf{n}_k\propto-\nabla m_k$, and apply a free-slip, no-tension contact projection:
\[
\tilde{\mathbf{v}}_k
=
\mathbf{v}_k
-
\max\!\left(
(\mathbf{v}_k-\mathbf{v}_\mathrm{cm})\cdot\mathbf{n}_k,
0
\right)\mathbf{n}_k .
\]
This projection removes only the normal component that drives the subject into contact, while preserving tangential motion and leaving separating motion unchanged. 
Consequently, approaching subjects are prevented from penetrating each other, whereas subjects that are already separating are not artificially constrained together. 
Nodes influenced by only a single subject are left unchanged, so the non-contact dynamics remain identical to the baseline MPM formulation.

As shown in Fig.~\ref{fig:artifact_alleviation}(c), this multi-field contact treatment mitigates the sticking between the object and the belly compared with Fig.~\ref{fig:artifact_alleviation}(b). 
Similarly, Fig.~\ref{fig:self_penetration}(b) shows reduced penetration between different body parts compared with Fig.~\ref{fig:self_penetration}(a).


\begin{figure*}[t]
\centering
\includegraphics[width=1.0\linewidth]{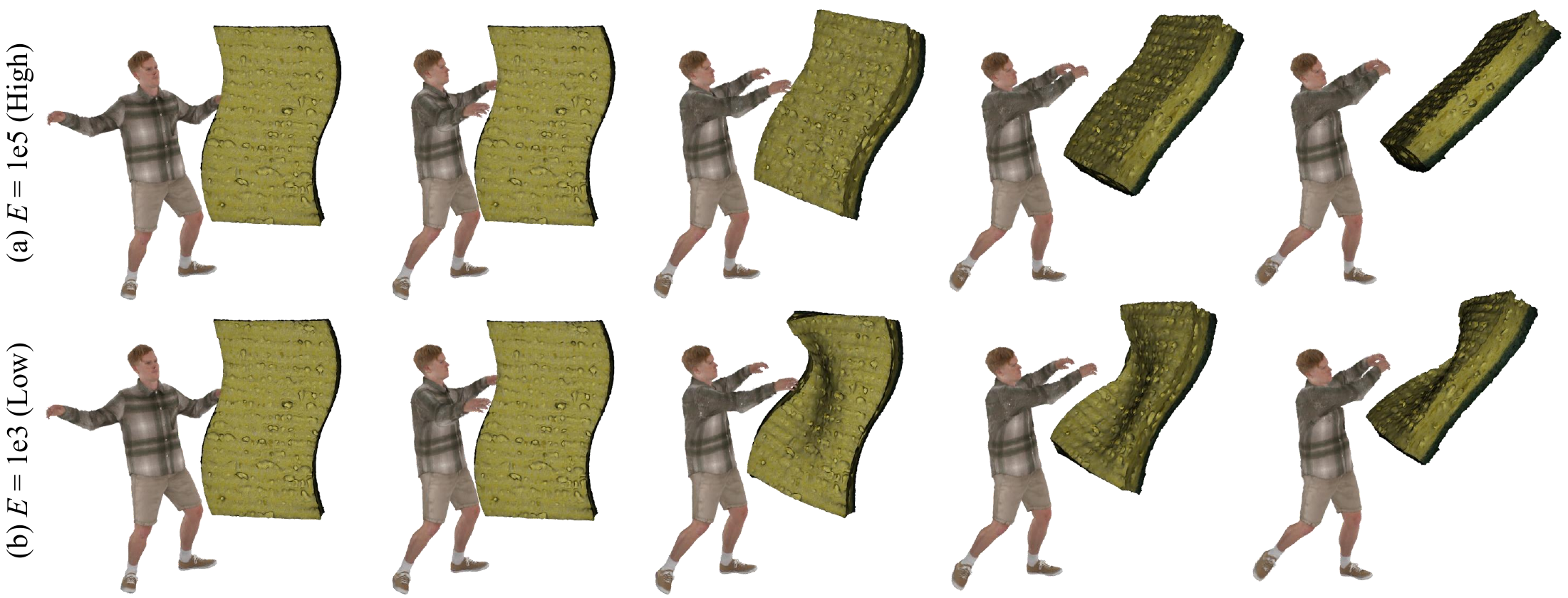}
\vspace{-10pt}
\caption{\textbf{Effect of Young's modulus $E$ on object stiffness.} (a) High-$E$ objects behave as rigid masses and move as solid blocks.
(b) Low-$E$ objects deform noticeably and fly in a soft, squishy manner.
These results demonstrate that our simulator naturally reflects material-dependent stiffness.}
\vspace{-3mm}
\label{fig:spongeE}
\end{figure*}

\begin{figure*}[t]
\centering
\includegraphics[width=1.0\linewidth]{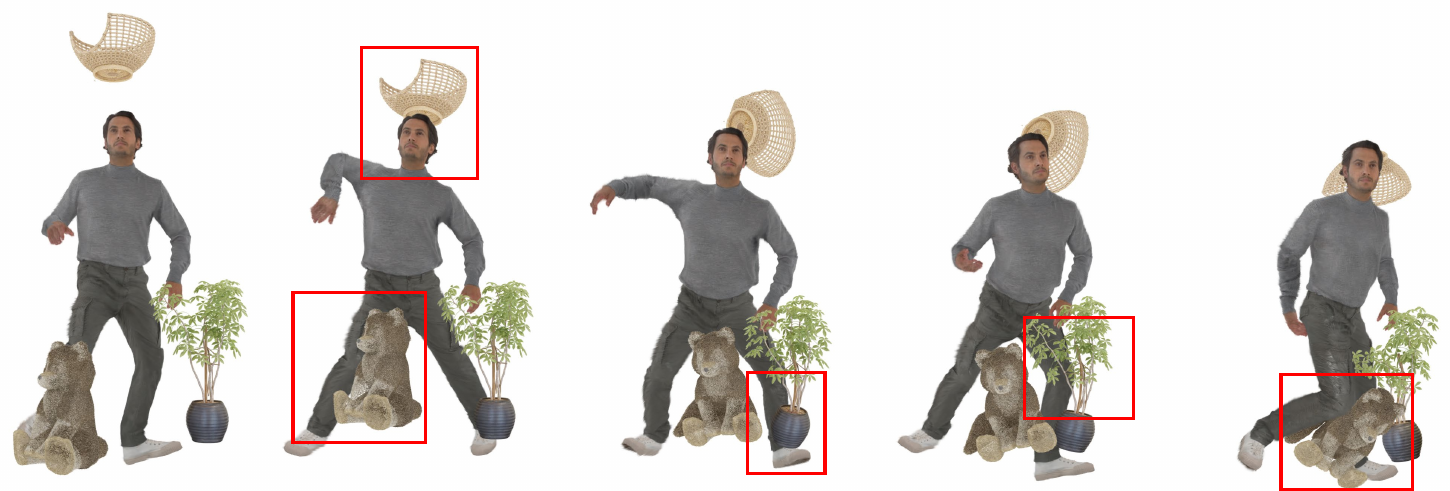}
\vspace{-10pt}
\caption{\textbf{Multi-object interactions.}
The avatar and objects push against each other, causing object deformation. Our simulator supports such interactions between an avatar and multiple dynamic objects, demonstrating multi-object capabilities.}
\label{fig:multi_object}
\end{figure*}

\begin{figure*}[t]
\centering
\includegraphics[width=1.0\linewidth]{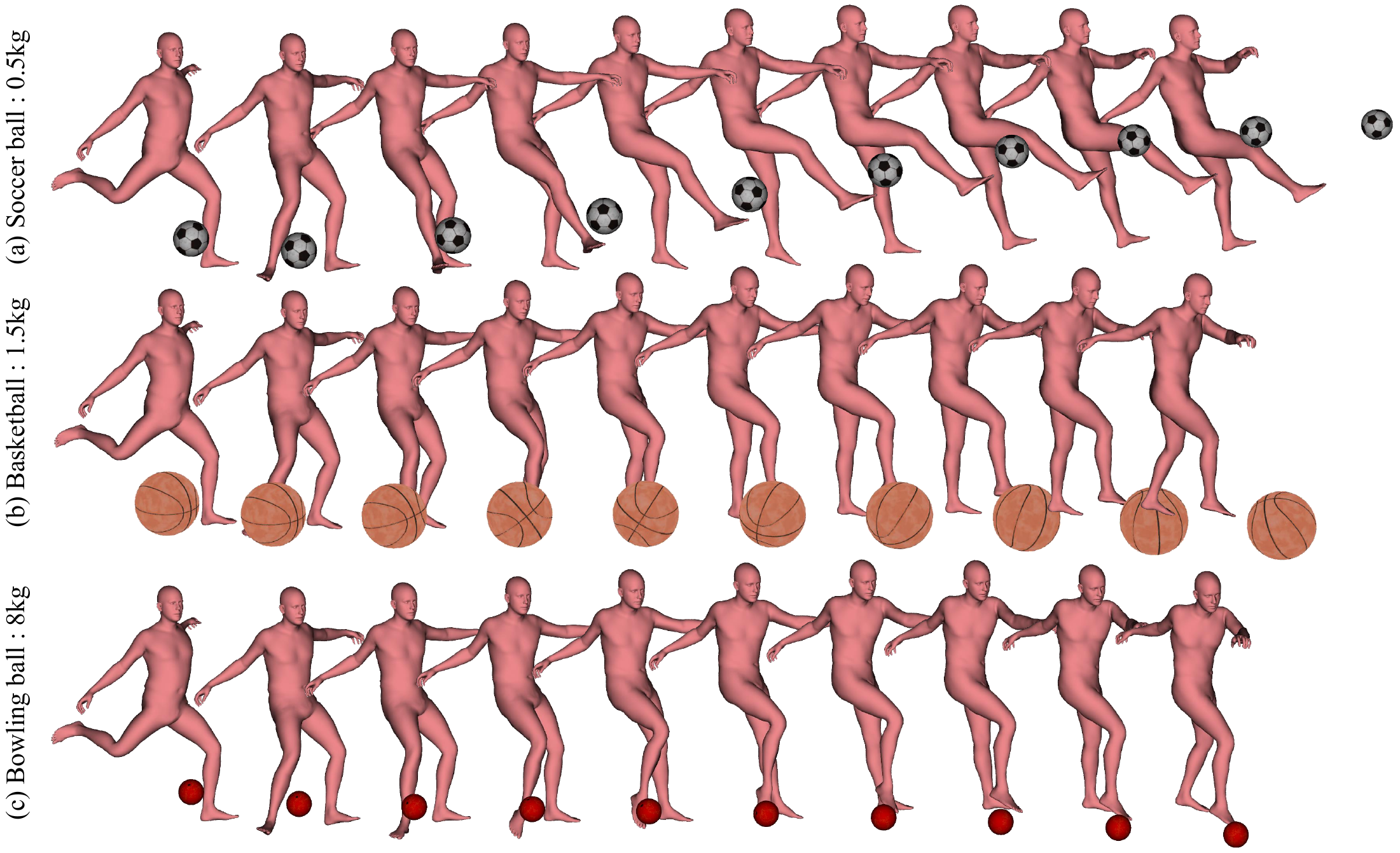}
\vspace{-10pt}
\caption{\textbf{Interaction response under different object masses.} We perform the same kicking motion with objects of different masses:
(a) soccer ball (0.5 kg), (b) basketball (1.5 kg), and (c) bowling ball (8 kg).
The simulation uses the same pose sequence and physical parameters for all cases, while only the object mass is changed.}
\vspace{-3mm}
\label{fig:kick_diff}
\end{figure*}

\begin{figure*}[t!]
    \centering
    \includegraphics[clip=true, width=\linewidth]{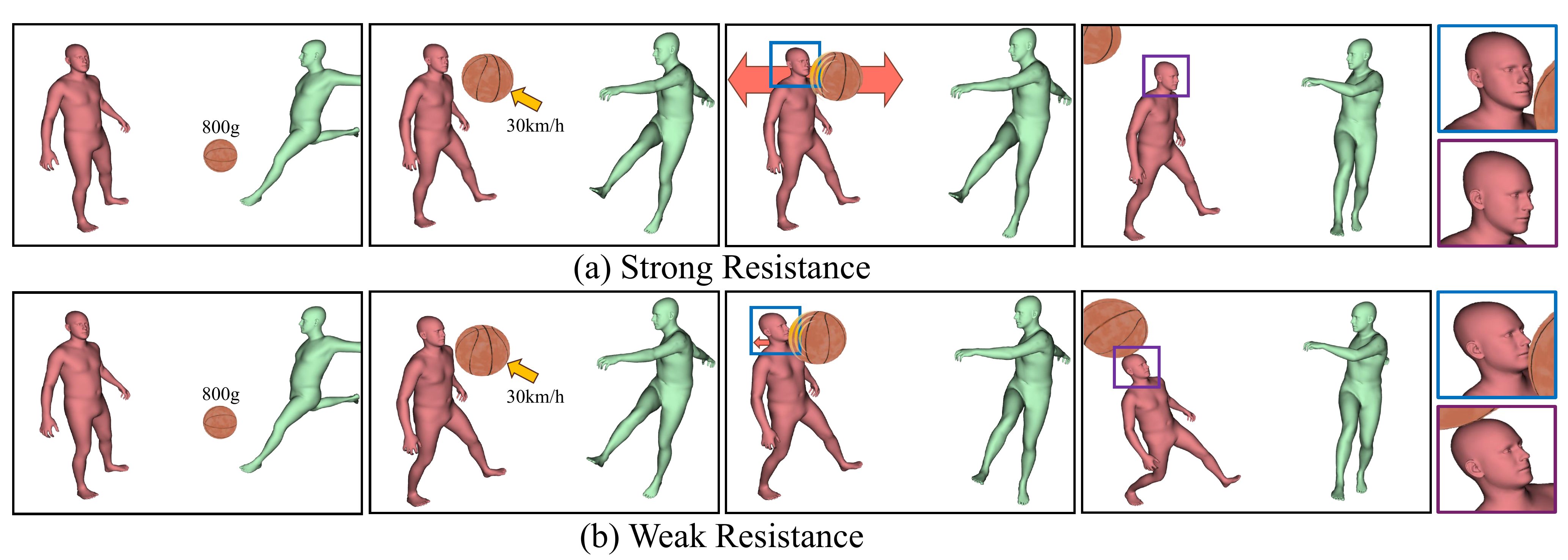}
    \vspace{-15pt}
    \caption{\textbf{Resistance simulation according to the impact of the ball.} The pink-colored avatar has the stronger resistance in (a) than in (b).}
    \vspace{-15pt}
    \label{fig:velo_diff}
\end{figure*}

\section{Additional Results}
We additionally present results on varying Young’s modulus, multi-object interactions, and skeleton particles.

\vspace{3pt}
\noindent\textbf{Young’s Modulus Variation.}
Fig.~\ref{fig:spongeE} shows how changes in Young’s modulus lead to different levels of stiffness, demonstrating that our simulator naturally reflects material properties.
Objects with high Young’s modulus, as in Fig.~\ref{fig:spongeE}(a), behave as rigid masses and travel as solid blocks, while those with low Young’s modulus, as in Fig.~\ref{fig:spongeE}(b), deform noticeably and move in a soft, squishy manner.

\vspace{3pt}
\noindent\textbf{Multi-object Interaction.} Fig.~\ref{fig:multi_object} illustrates interactions between multiple objects and an avatar; the avatar is pushed by objects approaching from the front, changes its posture in response to objects falling from above, and pushes other objects away. This indicates that our system also supports multi-object scenarios.

\vspace{3pt}
\noindent\textbf{Mass-dependent Interaction Response.}
We conduct an additional experiment in Fig.~\ref{fig:kick_diff}, similar to Fig.~\ref{fig:teaser}(a), while varying only the mass of the object.
The avatar follows the same pose sequence, but the result changes depending on the object inertia.
With a lighter object, the object motion changes significantly after contact while the avatar stays close to the target pose.
As the mass increases, stronger reaction forces cause larger pose deviation, whereas the object trajectory becomes less affected.
This result shows that the interaction behavior is governed by the object mass rather than the predefined motion.

\vspace{4pt}
\noindent\textbf{Simulation on Resistance of Avatar.}
To highlight the effectiveness of our PIAvatar, we design one more scenario where a ball, set in motion by one avatar's physical impact, strikes another avatar. In this setup, we simulate the realism of physical reactions by varying the resistive force of the struck avatar (pink-colored). As illustrated in Fig.~\ref{fig:velo_diff}, the avatar with high resistance only experiences a head recoil after being hit in the face (Fig.~\ref{fig:velo_diff}(a)), whereas the avatar with low resistance is simulated to collapse upon the impact (Fig.~\ref{fig:velo_diff}(b)).

\vspace{3pt}
\noindent\textbf{OSSO skeleton model particles.} 
The OSSO skeleton is composed of 74,496 particles. 
To address the overhead, we downsample them by 10$\times$ using voxel grid downsampling, reducing simulation time and Kabsch cost while maintaining comparable results (Tab.~\ref{tab:runtime_multi_rebuttal}).
Since the skeleton has an explicit shape, we do not place particles in empty space, and force transmission is preserved by sharing the grid between skeleton and body particles.

\begin{table}[t]
\caption{Runtime downsampled particles and Distance error.}
\vspace{-5pt}
\centering
\renewcommand{\arraystretch}{1.22}
\scriptsize
\setlength{\tabcolsep}{2pt}
\resizebox{0.95\linewidth}{!}{%
\begin{tabular}{l|cc|cc}
\hline
\noalign{\hrule height 1pt}
\textbf{Component} 
& \textbf{Single SMPL-X} 
& \textbf{Four SMPL-X} 
& \textbf{Single AG} 
& \textbf{Four AG} \\
\noalign{\hrule height 1pt}

\hline
Total (ms)\;$\downarrow$
& $1.745 \rightarrow \textbf{0.492}$
& $4.765 \rightarrow \textbf{1.436}$ 
& $4.385 \rightarrow \textbf{3.764}$ 
& $17.857 \rightarrow \textbf{15.259}$ \\
\cline{1-5}
Distance Error~(m)\;$\downarrow$         
& $0.0021 \rightarrow 0.0022 $
& $0.0019 \rightarrow 0.0021 $
& $0.0051 \rightarrow 0.0057 $
& $0.0054 \rightarrow 0.0052 $ \\

\noalign{\hrule height 1pt}
\hline
\end{tabular}}
\label{tab:runtime_multi_rebuttal}
\end{table}

%
%
\bibliographystyle{splncs04}
\bibliography{refs}

@string{CVPR = "Proceedings of IEEE Conference on Computer Vision and Pattern Recognition (CVPR)"}

@string{ICCV = "Proceedings of International Conference on Computer Vision (ICCV)"}

@string{ECCV = "Proceedings of European Conference on Computer Vision (ECCV)"}

@string{SIGGRAPH = "Proceedings of ACM SIGGRAPH"}

@string{PAMI = "IEEE Transactions on Pattern Analysis and Machine Intelligence (PAMI)"}

@string{TOG = "ACM Transactions on Graphics (TOG)"}

@string{ICLR = "International Conference on Learning Representations (ICLR)"}

@string{NIPS = "Proceedings of the Neural Information Processing Systems (NeurIPS)"}

@article{stomakhin2013material,
  title={A material point method for snow simulation},
  author={Stomakhin, Alexey and Schroeder, Craig and Chai, Lawrence and Teran, Joseph and Selle, Andrew},
  journal=TOG,
  year={2013},
  publisher={ACM New York, NY, USA}
}

@article{guo2018material,
  title={A material point method for thin shells with frictional contact},
  author={Guo, Qi and Han, Xuchen and Fu, Chuyuan and Gast, Theodore and Tamstorf, Rasmus and Teran, Joseph},
  journal=TOG,
  year={2018},
  publisher={ACM New York, NY, USA}
}

@article{han2019hybrid,
  title={A hybrid material point method for frictional contact with diverse materials},
  author={Han, Xuchen and Gast, Theodore F and Guo, Qi and Wang, Stephanie and Jiang, Chenfanfu and Teran, Joseph},
  journal=TOG,
  year={2019},
  publisher={ACM New York, NY, USA}
}

@article{jiang2017angular,
  title={An angular momentum conserving affine-particle-in-cell method},
  author={Jiang, Chenfanfu and Schroeder, Craig and Teran, Joseph},
  journal={Journal of Computational Physics (JCP)},
  year={2017},
  publisher={Elsevier}
}

@article{hu2018moving,
  title={A moving least squares material point method with displacement discontinuity and two-way rigid body coupling},
  author={Hu, Yuanming and Fang, Yu and Ge, Ziheng and Qu, Ziyin and Zhu, Yixin and Pradhana, Andre and Jiang, Chenfanfu},
  journal=TOG,
  year={2018},
  publisher={ACM New York, NY, USA}
}

@incollection{jiang2016material,
  title={The material point method for simulating continuum materials},
  author={Jiang, Chenfanfu and Schroeder, Craig and Teran, Joseph and Stomakhin, Alexey and Selle, Andrew},
  booktitle={Acm siggraph 2016 courses},
  year={2016},
  publisher = TOG,
}

@article{lee2018dexterous,
  title={Dexterous manipulation and control with volumetric muscles},
  author={Lee, Seunghwan and Yu, Ri and Park, Jungnam and Aanjaneya, Mridul and Sifakis, Eftychios and Lee, Jehee},
  journal=TOG,
  year={2018},
  publisher={ACM New York, NY, USA}
}

@inproceedings{keller2022osso,
  title={OSSO: Obtaining skeletal shape from outside},
  author={Keller, Marilyn and Zuffi, Silvia and Black, Michael J and Pujades, Sergi},
  booktitle=CVPR,
  year={2022}
}

@inproceedings{pavlakos2019expressive,
  title={Expressive body capture: 3d hands, face, and body from a single image},
  author={Pavlakos, Georgios and Choutas, Vasileios and Ghorbani, Nima and Bolkart, Timo and Osman, Ahmed AA and Tzionas, Dimitrios and Black, Michael J},
  booktitle=CVPR,
  year={2019}
}

@inproceedings{li2024animatable,
  title={Animatable gaussians: Learning pose-dependent gaussian maps for high-fidelity human avatar modeling},
  author={Li, Zhe and Zheng, Zerong and Wang, Lizhen and Liu, Yebin},
  booktitle=CVPR,
  year={2024}
}

@inproceedings{hu2024gaussianavatar,
  title={Gaussianavatar: Towards realistic human avatar modeling from a single video via animatable 3d gaussians},
  author={Hu, Liangxiao and Zhang, Hongwen and Zhang, Yuxiang and Zhou, Boyao and Liu, Boning and Zhang, Shengping and Nie, Liqiang},
  booktitle=CVPR,
  year={2024}
}

@article{xue2024human,
  title={Human 3Diffusion: Realistic Avatar Creation via Explicit 3D Consistent Diffusion Models},
  author={Xue, Yuxuan and Xie, Xianghui and Marin, Riccardo and Pons-Moll, Gerard},
  journal=NIPS,
  year={2024}
}

@article{kerbl20233d,
  title={3D Gaussian Splatting for Real-Time Radiance Field Rendering.},
  author={Kerbl, Bernhard and Kopanas, Georgios and Leimk{\"u}hler, Thomas and Drettakis, George},
  journal=TOG,
  year={2023}
}

@inproceedings{saito2019pifu,
  title={Pifu: Pixel-aligned implicit function for high-resolution clothed human digitization},
  author={Saito, Shunsuke and Huang, Zeng and Natsume, Ryota and Morishima, Shigeo and Kanazawa, Angjoo and Li, Hao},
  booktitle=ICCV,
  year={2019}
}

@inproceedings{saito2020pifuhd,
  title={Pifuhd: Multi-level pixel-aligned implicit function for high-resolution 3d human digitization},
  author={Saito, Shunsuke and Simon, Tomas and Saragih, Jason and Joo, Hanbyul},
  booktitle=CVPR,
  year={2020}
}

@inproceedings{xiu2023econ,
  title={Econ: Explicit clothed humans optimized via normal integration},
  author={Xiu, Yuliang and Yang, Jinlong and Cao, Xu and Tzionas, Dimitrios and Black, Michael J},
  booktitle=CVPR,
  year={2023}
}

@inproceedings{xiu2022icon,
  title={Icon: Implicit clothed humans obtained from normals},
  author={Xiu, Yuliang and Yang, Jinlong and Tzionas, Dimitrios and Black, Michael J},
  booktitle=CVPR,
  year={2022},
}

@article{zheng2021pamir,
  title={Pamir: Parametric model-conditioned implicit representation for image-based human reconstruction},
  author={Zheng, Zerong and Yu, Tao and Liu, Yebin and Dai, Qionghai},
  journal=PAMI,
  year={2021},
  publisher={IEEE}
}

@inproceedings{han2023high,
  title={High-fidelity 3d human digitization from single 2k resolution images},
  author={Han, Sang-Hun and Park, Min-Gyu and Yoon, Ju Hong and Kang, Ju-Mi and Park, Young-Jae and Jeon, Hae-Gon},
  booktitle=CVPR,
  year={2023}
}

@inproceedings{ho2024sith,
  title={Sith: Single-view textured human reconstruction with image-conditioned diffusion},
  author={Ho, I and Song, Jie and Hilliges, Otmar and others},
  booktitle=CVPR,
  year={2024}
}

@article{isik2023humanrf,
  title = {HumanRF: High-Fidelity Neural Radiance Fields for Humans in Motion},
  author = {I\c{s}{\i}k, Mustafa and Rünz, Martin and Georgopoulos, Markos and Khakhulin, Taras
    and Starck, Jonathan and Agapito, Lourdes and Nießner, Matthias},
  journal = TOG,
  year = {2023},
  publisher = {ACM New York, NY, USA},
  doi = {10.1145/3592415},
  url = {https://doi.org/10.1145/3592415}
}

@inproceedings{peng2021neural,
  title={Neural body: Implicit neural representations with structured latent codes for novel view synthesis of dynamic humans},
  author={Peng, Sida and Zhang, Yuanqing and Xu, Yinghao and Wang, Qianqian and Shuai, Qing and Bao, Hujun and Zhou, Xiaowei},
  booktitle=CVPR,
  year={2021}
}

@inproceedings{weng2022humannerf,
  title={Humannerf: Free-viewpoint rendering of moving people from monocular video},
  author={Weng, Chung-Yi and Curless, Brian and Srinivasan, Pratul P and Barron, Jonathan T and Kemelmacher-Shlizerman, Ira},
  booktitle=CVPR,
  year={2022}
}

@article{su2021nerf,
  title={A-nerf: Articulated neural radiance fields for learning human shape, appearance, and pose},
  author={Su, Shih-Yang and Yu, Frank and Zollh{\"o}fer, Michael and Rhodin, Helge},
  journal=NIPS,
  year={2021}
}

@inproceedings{peng2021animatable,
  title={Animatable neural radiance fields for modeling dynamic human bodies},
  author={Peng, Sida and Dong, Junting and Wang, Qianqian and Zhang, Shangzhan and Shuai, Qing and Zhou, Xiaowei and Bao, Hujun},
  booktitle=ICCV,
  year={2021}
}

@inproceedings{guo2023vid2avatar,
  title={Vid2avatar: 3d avatar reconstruction from videos in the wild via self-supervised scene decomposition},
  author={Guo, Chen and Jiang, Tianjian and Chen, Xu and Song, Jie and Hilliges, Otmar},
  booktitle=CVPR,
  year={2023}
}

@inproceedings{jiang2023instantavatar,
  title={Instantavatar: Learning avatars from monocular video in 60 seconds},
  author={Jiang, Tianjian and Chen, Xu and Song, Jie and Hilliges, Otmar},
  booktitle=CVPR,
  year={2023}
}

@inproceedings{wen2024gomavatar,
  title={Gomavatar: Efficient animatable human modeling from monocular video using gaussians-on-mesh},
  author={Wen, Jing and Zhao, Xiaoming and Ren, Zhongzheng and Schwing, Alexander G and Wang, Shenlong},
  booktitle=CVPR,
  year={2024}
}

@inproceedings{shin2024canonicalfusion,
  title={Canonicalfusion: Generating drivable 3d human avatars from multiple images},
  author={Shin, Jisu and Lee, Junmyeong and Lee, Seongmin and Park, Min-Gyu and Kang, Ju-Mi and Yoon, Ju Hong and Jeon, Hae-Gon},
  booktitle=ECCV,
  year={2024},
}

@article{pan2024humansplat,
  title={Humansplat: Generalizable single-image human gaussian splatting with structure priors},
  author={Pan, Panwang and Su, Zhuo and Lin, Chenguo and Fan, Zhen and Zhang, Yongjie and Li, Zeming and Shen, Tingting and Mu, Yadong and Liu, Yebin},
  journal=NIPS,
  year={2024}
}

@inproceedings{moon2024expressive,
  title={Expressive whole-body 3d gaussian avatar},
  author={Moon, Gyeongsik and Shiratori, Takaaki and Saito, Shunsuke},
  booktitle=ECCV,
  year={2024},
}

@article{keller2023skin,
  title={From skin to skeleton: Towards biomechanically accurate 3d digital humans},
  author={Keller, Marilyn and Werling, Keenon and Shin, Soyong and Delp, Scott and Pujades, Sergi and Liu, C Karen and Black, Michael J},
  journal=TOG,
  year={2023},
  publisher={ACM New York, NY, USA}
}

@article{loper2015smpl,
    author = {Loper, Matthew and Mahmood, Naureen and Romero, Javier and Gerard Pons-Moll and Black, Michael J.},
    title = {SMPL: a skinned multi-person linear model},
    journal = TOG,
    year = {2015},
}

@inproceedings{SMPLX_2019_Pavlakos,
  author = {Pavlakos, Georgios and Choutas, Vasileios and Ghorbani, Nima and Bolkart, Timo and Osman, Ahmed A. A. and Tzionas, Dimitrios and Black, Michael J.},
  title = {Expressive Body Capture: {3D} Hands, Face, and Body from a Single Image},
  booktitle = CVPR,
  year = {2019}
}

@incollection{anguelov2005scape,
  title={Scape: shape completion and animation of people},
  author={Anguelov, Dragomir and Srinivasan, Praveen and Koller, Daphne and Thrun, Sebastian and Rodgers, Jim and Davis, James},
  booktitle=TOG,
  year={2005},
  publisher = {ACM}
}

@inproceedings{xu2020ghum,
  title={Ghum \& ghuml: Generative 3d human shape and articulated pose models},
  author={Xu, Hongyi and Bazavan, Eduard Gabriel and Zanfir, Andrei and Freeman, William T and Sukthankar, Rahul and Sminchisescu, Cristian},
  booktitle=CVPR,
  year={2020}
}

@inproceedings{Bogo:ECCV:2016,
title = {Keep it {SMPL}: Automatic Estimation of {3D} Human Pose and Shape
from a Single Image},
author = {Bogo, Federica and Kanazawa, Angjoo and Lassner, Christoph and
Gehler, Peter and Romero, Javier and Black, Michael J.},
booktitle = ECCV,
year = {2016}
}

@article{loper2014mosh,
  title={MoSh: motion and shape capture from sparse markers.},
  author={Loper, Matthew and Mahmood, Naureen and Black, Michael J},
  journal=TOG,
  year={2014}
}

@InProceedings{hmrKanazawa17,
  title={End-to-end Recovery of Human Shape and Pose},
  author = {Angjoo Kanazawa and Michael J. Black and David W. Jacobs and Jitendra Malik},
  booktitle=CVPR,
  year={2018}
}

@inproceedings{kocabas2019vibe,
  title={VIBE: Video Inference for Human Body Pose and Shape Estimation},
  author={Kocabas, Muhammed and Athanasiou, Nikos and Black, Michael J.},
  booktitle =CVPR,
  year = {2020}
}

@misc{maximeraafat_BlenderNeRF,
  author       = {maximeraafat},
  title        = {BlenderNeRF},
  year         = {2023},
  publisher    = {GitHub},
  howpublished = {\url{https://github.com/maximeraafat/BlenderNeRF}}  
}

@inproceedings{mahmood2019amass,
  title={AMASS: Archive of motion capture as surface shapes},
  author={Mahmood, Naureen and Ghorbani, Nima and Troje, Nikolaus F and Pons-Moll, Gerard and Black, Michael J},
  booktitle=ICCV,
  year={2019}
}

@misc{Genesis,
          author = {Genesis Authors},
          title = {Genesis: A Generative and Universal Physics Engine for Robotics and Beyond},
          month = {December},
          year = {2024},
          url = {https://github.com/Genesis-Embodied-AI/Genesis}
        }

@article{
	2018-TOG-deepMimic,
	author = {Peng, Xue Bin and Abbeel, Pieter and Levine, Sergey and van de Panne, Michiel},
	title = {DeepMimic: Example-guided Deep Reinforcement Learning of Physics-based Character Skills},
	journal = TOG,
	year = {2018},
}

@incollection{coumans2015bullet,
  title={Bullet physics simulation},
  author={Coumans, Erwin},
  booktitle={ACM SIGGRAPH 2015 Courses},
  pages={1},
  year={2015},
  publisher = {ACM}
}

@misc{nvidiaPhysX,
  title        = {NVIDIA PhysX SDK},
  author       = {NVIDIA Corporation},
  publisher    = {NVIDIA Corporation},
  year         = {2021},
  howpublished = {\url{https://developer.nvidia.com/physx-sdk}},
  note         = {Accessed: March 2025}
}

@inproceedings{todorov2012mujoco,
  title={Mujoco: A physics engine for model-based control},
  author={Todorov, Emanuel and Erez, Tom and Tassa, Yuval},
  booktitle={Proceedings of International Conference on Intelligent Robots and Systems (IROS)},
  year={2012},
  organization={IEEE}
}

@inproceedings{xie2024physgaussian,
  title={Physgaussian: Physics-integrated 3d gaussians for generative dynamics},
  author={Xie, Tianyi and Zong, Zeshun and Qiu, Yuxing and Li, Xuan and Feng, Yutao and Yang, Yin and Jiang, Chenfanfu},
  booktitle=CVPR,
  year={2024}
}

@article{kabsch1976solution,
  title={A solution for the best rotation to relate two sets of vectors},
  author={Kabsch, Wolfgang},
  journal={Foundations of Crystallography},
  year={1976},
  publisher={International Union of Crystallography}
}

@article{jiang2015affine,
  title={The affine particle-in-cell method},
  author={Jiang, Chenfanfu and Schroeder, Craig and Selle, Andrew and Teran, Joseph and Stomakhin, Alexey},
  journal=TOG,
  year={2015},
  publisher={ACM New York, NY, USA}
}

@article{klar2016drucker,
  title={Drucker-prager elastoplasticity for sand animation},
  author={Kl{\'a}r, Gergely and Gast, Theodore and Pradhana, Andre and Fu, Chuyuan and Schroeder, Craig and Jiang, Chenfanfu and Teran, Joseph},
  journal=TOG,
  year={2016},
  publisher={ACM New York, NY, USA}
}

@article{jiang2017anisotropic,
  title={Anisotropic elastoplasticity for cloth, knit and hair frictional contact},
  author={Jiang, Chenfanfu and Gast, Theodore and Teran, Joseph},
  journal=TOG,
  year={2017},
  publisher={ACM New York, NY, USA}
}

@article{tevet2024closd,
  title={Closd: Closing the loop between simulation and diffusion for multi-task character control},
  author={Tevet, Guy and Raab, Sigal and Cohan, Setareh and Reda, Daniele and Luo, Zhengyi and Peng, Xue Bin and Bermano, Amit H and van de Panne, Michiel},
  journal=ICLR,
  year={2025}
}

@inproceedings{xu2025intermimic,
  title={Intermimic: Towards universal whole-body control for physics-based human-object interactions},
  author={Xu, Sirui and Ling, Hung Yu and Wang, Yu-Xiong and Gui, Liang-Yan},
  booktitle=CVPR,
  year={2025}
}

@inproceedings{luo2023perpetual,
  title={Perpetual humanoid control for real-time simulated avatars},
  author={Luo, Zhengyi and Cao, Jinkun and Kitani, Kris and Xu, Weipeng and others},
  booktitle=ICCV,
  year={2023}
}

@article{peng2021amp,
  title={Amp: Adversarial motion priors for stylized physics-based character control},
  author={Peng, Xue Bin and Ma, Ze and Abbeel, Pieter and Levine, Sergey and Kanazawa, Angjoo},
  journal=TOG,
  year={2021},
  publisher={ACM New York, NY, USA}
}

@article{luo2023universal,
  title={Universal humanoid motion representations for physics-based control},
  author={Luo, Zhengyi and Cao, Jinkun and Merel, Josh and Winkler, Alexander and Huang, Jing and Kitani, Kris and Xu, Weipeng},
  journal=ICLR,
  year={2023}
}

@inproceedings{zheng2024physavatar,
  title={Physavatar: Learning the physics of dressed 3d avatars from visual observations},
  author={Zheng, Yang and Zhao, Qingqing and Yang, Guandao and Yifan, Wang and Xiang, Donglai and Dubost, Florian and Lagun, Dmitry and Beeler, Thabo and Tombari, Federico and Guibas, Leonidas and others},
  booktitle=ECCV,
  year={2024}
}

@inproceedings{lee2025mpmavatar,
  title={Mpmavatar: Learning 3d gaussian avatars with accurate and robust physics-based dynamics},
  author={Lee, Changmin and Lee, Jihyun and Kim, Tae-Kyun},
  booktitle=NIPS,
  year={2025}
}

@article{siyao2025half,
  title={Half-Physics: Enabling Kinematic 3D Human Model with Physical Interactions},
  author={Siyao, Li and Feng, Yao and Taheri, Omid and Loy, Chen Change and Black, Michael J},
  journal={arXiv preprint arXiv:2507.23778},
  year={2025}
}

@article{muller2005meshless,
  title={Meshless deformations based on shape matching},
  author={M{\"u}ller, Matthias and Heidelberger, Bruno and Teschner, Matthias and Gross, Markus},
  journal={ACM transactions on graphics (TOG)},
  volume={24},
  number={3},
  pages={471--478},
  year={2005},
  publisher={ACM New York, NY, USA}
}

@article{loccoz20243dgrt,
    author = {Nicolas Moenne-Loccoz and Ashkan Mirzaei and Or Perel and Riccardo de Lutio and Janick Martinez Esturo and Gavriel State and Sanja Fidler and Nicholas Sharp and Zan Gojcic},
    title = {3D Gaussian Ray Tracing: Fast Tracing of Particle Scenes},
    journal = TOG,
    year = {2024},
}

@article{wu20253dgut,
    title={3DGUT: Enabling Distorted Cameras and Secondary Rays in Gaussian Splatting},
    author={Wu, Qi and Martinez Esturo, Janick and Mirzaei, Ashkan and Moenne-Loccoz, Nicolas and Gojcic, Zan},
    journal = {Conference on Computer Vision and Pattern Recognition (CVPR)},
    year={2025}
}

@article{herlihy1991wait,
  title={Wait-free synchronization},
  author={Herlihy, Maurice},
  journal={ACM Transactions on Programming Languages and Systems (TOPLAS)},
  year={1991},
  publisher={ACM New York, NY, USA}
}

@article{bardenhagen2000material,
  title={The Material-Point Method for Granular Materials},
  author={Bardenhagen, S. G. and Brackbill, J. U. and Sulsky, D.},
  journal={Computer Methods in Applied Mechanics and Engineering},
  volume={187},
  number={3--4},
  pages={529--541},
  year={2000}
}

@inproceedings{shao2024splattingavatar,
  title={Splattingavatar: Realistic real-time human avatars with mesh-embedded gaussian splatting},
  author={Shao, Zhijing and Wang, Zhaolong and Li, Zhuang and Wang, Duotun and Lin, Xiangru and Zhang, Yu and Fan, Mingming and Wang, Zeyu},
  booktitle=CVPR,
  year={2024}
}

@article{hu2024expressive,
  title={Expressive gaussian human avatars from monocular rgb video},
  author={Hu, Hezhen and Fan, Zhiwen and Wu, Tianhao and Xi, Yihan and Lee, Seoyoung and Pavlakos, Georgios and Wang, Zhangyang and others},
  journal=NIPS,
  year={2024}
}

@inproceedings{zhuang2025idol,
  title={Idol: Instant photorealistic 3d human creation from a single image},
  author={Zhuang, Yiyu and Lv, Jiaxi and Wen, Hao and Shuai, Qing and Zeng, Ailing and Zhu, Hao and Chen, Shifeng and Yang, Yujiu and Cao, Xun and Liu, Wei},
  booktitle=CVPR,
  year={2025}
}

@inproceedings{sim2025persona,
  title={PERSONA: Personalized Whole-Body 3D Avatar with Pose-Driven Deformations from a Single Image},
  author={Sim, Geonhee and Moon, Gyeongsik},
  booktitle=ICCV,
  year={2025}
}

@article{qiu2025lhm,
  title={Lhm: Large animatable human reconstruction model from a single image in seconds},
  author={Qiu, Lingteng and Gu, Xiaodong and Li, Peihao and Zuo, Qi and Shen, Weichao and Zhang, Junfei and Qiu, Kejie and Yuan, Weihao and Chen, Guanying and Dong, Zilong and others},
  journal=ICCV,
  year={2025}
}

@inproceedings{qiu2025anigs,
  title={Anigs: Animatable gaussian avatar from a single image with inconsistent gaussian reconstruction},
  author={Qiu, Lingteng and Zhu, Shenhao and Zuo, Qi and Gu, Xiaodong and Dong, Yuan and Zhang, Junfei and Xu, Chao and Li, Zhe and Yuan, Weihao and Bo, Liefeng and others},
  booktitle=CVPR,
  year={2025}
}

@inproceedings{wang2025fresa,
  title={FRESA: Feedforward Reconstruction of Personalized Skinned Avatars from Few Images},
  author={Wang, Rong and Prada, Fabian and Wang, Ziyan and Jiang, Zhongshi and Yin, Chengxiang and Li, Junxuan and Saito, Shunsuke and Santesteban, Igor and Romero, Javier and Joshi, Rohan and others},
  booktitle=CVPR,
  year={2025}
}

@inproceedings{wu20244d,
  title={4d gaussian splatting for real-time dynamic scene rendering},
  author={Wu, Guanjun and Yi, Taoran and Fang, Jiemin and Xie, Lingxi and Zhang, Xiaopeng and Wei, Wei and Liu, Wenyu and Tian, Qi and Wang, Xinggang},
  booktitle=CVPR,
  year={2024}
}

@article{liu2013simulation,
  title={Simulation and control of skeleton-driven soft body characters},
  author={Liu, Libin and Yin, KangKang and Wang, Bin and Guo, Baining},
  journal={ACM Transactions on Graphics (TOG)},
  year={2013},
  publisher={ACM New York, NY, USA}
}

@article{kim2017data,
  title={Data-driven physics for human soft tissue animation},
  author={Kim, Meekyoung and Pons-Moll, Gerard and Pujades, Sergi and Bang, Seungbae and Kim, Jinwook and Black, Michael J and Lee, Sung-Hee},
  journal={ACM Transactions on Graphics (TOG)},
  year={2017},
  publisher={ACM New York, NY, USA}
}

@article{li2020codimensional,
  title={Codimensional incremental potential contact},
  author={Li, Minchen and Kaufman, Danny M and Jiang, Chenfanfu},
  journal={arXiv preprint arXiv:2012.04457},
  year={2020}
}

@book{clough1960finite,
  title={The Finite Element Method in Plane Stress Analysis},
  author={Clough, R.W.},
  url={https://books.google.co.kr/books?id=rfwFHQAACAAJ},
  year={1960},
  publisher={American Society of Civil Engineers}
}

@article{Qwen2.5-VL,
  title={Qwen2.5-VL Technical Report},
  author={Bai, Shuai and Chen, Keqin and Liu, Xuejing and Wang, Jialin and Ge, Wenbin and Song, Sibo and Dang, Kai and Wang, Peng and Wang, Shijie and Tang, Jun and Zhong, Humen and Zhu, Yuanzhi and Yang, Mingkun and Li, Zhaohai and Wan, Jianqiang and Wang, Pengfei and Ding, Wei and Fu, Zheren and Xu, Yiheng and Ye, Jiabo and Zhang, Xi and Xie, Tianbao and Cheng, Zesen and Zhang, Hang and Yang, Zhibo and Xu, Haiyang and Lin, Junyang},
  journal={arXiv preprint arXiv:2502.13923},
  year={2025}
}

@incollection{frazier2018bayesian,
  title={Bayesian optimization},
  author={Frazier, Peter I},
  booktitle={Recent advances in optimization and modeling of contemporary problems},
  year={2018},
  publisher={Informs}
}

\end{document}